\documentclass[aps,twocolumn,prx,showpacs,superscriptaddress,10pt]{revtex4-2}

\usepackage{amsmath, amsthm, amssymb, nicefrac,mathtools,dsfont,bm,bbm, braket, relsize, enumitem} 
\usepackage[dvipsnames]{xcolor}  
\usepackage{lmodern,microtype,ragged2e, comment, appendix, hyphenat} 
\usepackage{graphicx}
\usepackage{dcolumn}
\definecolor{mygrey}{gray}{0.35}
\definecolor{myblue}{rgb}{0.2,0.2,0.8}
\definecolor{myzard}{cmyk}{0,0,0.05,0}
\definecolor{mywhite}{rgb}{1,1,1}
\definecolor{myred}{rgb}{0.9,0.1,0.}
\definecolor{dgreen}{rgb}{0.0, 0.5, 0.0}
\usepackage[colorlinks=true,citecolor=myblue,linkcolor=myblue,urlcolor=myblue]{hyperref}
\usepackage[capitalize]{cleveref}
\usepackage[normalem]{ulem} 

\newcommand{\proj}[1]{\ket{#1}\!\bra{#1}}
\newcommand{\ketbra}[2]{\left|#1\middle\rangle\!\middle\langle#2\right|}

\newcommand{\inner}[2]{\left\langle #1 \middle\vert #2 \right\rangle}

\newcommand{\abs}[1]{\left\lvert #1\right\rvert}

\newcommand{\be}{\begin{equation}} 							
\newcommand{\ee}{\end{equation}}
\newcommand{\bematrix}{\left(\begin{matrix}}
\newcommand{\ematrix}{\end{matrix}\right)}
\def\one{{\mbox{$1 \hspace{-1.0mm}  {\bf l}$}}}						
\def\C{{\ensuremath{\mathbbm C}}}
\def\R{{\ensuremath{\mathbbm R}}}

\def\ii{\mathrm{i}}
\def\tr{\mathrm{tr}}
\def\d{\mathrm{d}}

\def\paulivec{\bm \sigma}

\def\cF{\mathcal F}

\def\cH{\mathcal H}

\def\cO{\mathcal O}

\newcommand{\eqnref}[1]{Eq.~(\ref{#1})}
\newcommand{\eqnsref}[2]{Eqs.~(\ref{#1}) and (\ref{#2})}
\newcommand{\figref}[1]{Fig.~\ref{#1}}
\newcommand{\figsref}[2]{Figs.~\ref{#1} and \ref{#2}}

\newcommand{\secref}[1]{Sec.~\ref{#1}}

\begin{document}
\title{Super-resolution of partially coherent bosonic sources}
\date{\today}
\author{Joaqu\'in L\'opez-Su\'arez}\email[Corresponding author:]{joaquinlopezsuarez@gmail.com}
\affiliation{Quantum Thermodynamics and Computation Group, Department of Electromagnetism and Condensed Matter, 
Universidad de Granada, 18071 Granada, Spain\looseness=-1}
\affiliation{Department of Physics, Stockholm University, AlbaNova University Center, SE-106 91 Stockholm, Sweden\looseness=-1}
\affiliation{Department of Nuclear and Particle Physics, Faculty of Physics, National and Kapodistrian University of Athens, 15784 Athens, Greece}
\author{Michalis Skotiniotis}\email[Corresponding author:]{mskotiniotis@onsager.ugr.es}
\affiliation{Quantum Thermodynamics and Computation Group, Department of Electromagnetism and Condensed Matter, 
Universidad de Granada, 18071 Granada, Spain\looseness=-1}
\affiliation{Instituto Carlos I de Física Teórica y Computacional, Universidad de Granada, 18071 Granada, Spain\looseness=-1}

\begin{abstract}
We consider imaging of two partially coherent sources and derive the ultimate quantum limits for estimating the separation, location, 
relative intensity, and coherence factor. We show that super-resolution in the separation is achievable both in the presence 
of nuisance parameters as well as when some of the parameters are assumed known. Nuisance parameters restrict super-resolution to 
balanced sources, whereas for known parameters super-resolution persists over a broader range of relative intensities and is lost 
only for perfectly correlated sources, i.e., $\gamma=1$. The achievable precision is governed primarily by 
interference-induced photon statistics and depends strongly on the degree of coherence. In the sub-Rayleigh regime, the imaging 
problem reduces to an effective two-dimensional Hilbert space description, provided a consistent reference position is used, with 
all parameters encoded in a Bloch vector representation. Finally, indirect estimation schemes based on the purity of the image-plane 
state are generally suboptimal for all non-zero coherence and become valid only in the incoherent limit.
\end{abstract}

\maketitle

\section{\label{sec:Introduction}Introduction}

The resolution of conventional imaging systems is known to be limited by the diffraction properties of the electromagnetic field~\cite{Rayleigh80}. 
Recently, techniques from quantum statistical inference have shown that the diffraction limit can be entirely circumvented for imaging of incoherent point 
sources~\cite{Tsang16,Tsang16b,Nair16,Lupo16,Ang17,Rehacek17,Rehacek18,Lu18,Napoli19,Prasad19,Tsang19,Lupo20,deAlmeida21,Hu25,Katamadze25},
and have been corroborated by experiments~\cite{Yang16,Tham17,Yang17,Donohue18,Parniak18,Zhou19,Boucher20,Rouviere24}.
The debate as to whether the same phenomenon can be exhibited for partially coherent sources~\cite{Larson2018,Tsang2019b,Larson2019b} was resolved 
in favour of super-resolution~\cite{Hradil2019,Hradil2021,Liang2021,Karuseichyk2022} and vindicated by experiments~\cite{Wadood21}. Since then, 
super-resolution in the presence of partial coherence has been shown to be possible for thermal sources~\cite{Karuseichyk2024}, as well as for Gaussian 
states~\cite{Sorelli2022}.

In this work we consider the full quantum mechanical treatment for estimating all pertinent parameters in the imaging of two 
partially coherent point-like sources: these are the separation $s>0$, a location parameter $x_0$, relative intensity $0\leq q \leq 1$, and coherence 
factor $\gamma\in\mathbb{C}, \,|\gamma|\leq 1$. We derive the quantum measurements that maximize the precision in 
estimating each parameter of interest and determine their compatibility. We study the
estimation precision for each of the pertinent parameters both in the case where some parameters are known or are treated
as nuisance parameters.  Our results can be summarized as follows:
	\begin{enumerate}
		\item
		
		Super-resolution in the estimation of the separation between two sources is achievable both when all parameters are known 
		and in the presence of nuisance parameters, albeit with strictly reduced precision in the latter case. This is contrary to the findings 
		reported in~\cite{Larson2018, Liang2021}. Moreover, the range of relative intensities $q$ for which super-resolution is 
        attainable is broader when all parameters are known, and is lost only  in the limiting case when $\gamma=1$ and $q=1/2$.
		
		\item
		
		We show that the photon counting statistics per coherence time interval provide the dominant contribution to the precision in estimating 
		the separation. Furthermore, we demonstrate that previous claims of a resurgence of Rayleigh’s curse~\cite{Larson2018, Liang2021} arise from 
		neglecting this information.
		
		\item
		
		The precision in estimating the source separation depends sensitively on the reference position $x_0$ used to define the Point Spread Functions
		(PSFs). Treating $x_0$ as known yields strictly higher precision than when it is treated as a nuisance parameter. When $x_0$ coincides with the 
		geometric centre of the two PSFs, the precision achieves its maximal value. 
		
		\item 
		
		We identify the conditions under which the imaging of two partially coherent sources can be reduced to an effective two-dimensional 
		Hilbert space description, greatly simplifying the analysis~\cite{Chrostowski2017}. This reduction is faithful provided that both the 
		PSFs and the measurement modes are defined with respect to the same reference position $x_0$, assumed to be known.
		
		\item 
		
		Indirect estimation schemes that infer the source separation from the purity of the density operator at the image plane are, 
		in general, suboptimal. Moreover, we show that they can fail when the real part of the coherence factor is negative. The only 
		regime in which such approaches remain valid is the imaging of incoherent sources, corresponding to $\gamma=0$.
		  
	\end{enumerate}

The paper is structured as follows.  \secref{sec:Background} reviews imaging in both the classical 
(\secref{sec:classical_imaging}) and quantum (\secref{sec:quantum_imaging}) regimes, the qubit 
model of~\cite{Chrostowski2017} (\secref{sec:qubit_model}), as well as the necessary background on both 
classical and quantum statistical inference (\secref{subs:param_est}).  In \secref{sec:qfi} we compute the 
precisions for each of the 4 parameters in question with (\secref{sec:qubit_model_validity}) 
and without (\secref{subs:qfi_direct}) the qubit model approximation, whilst in \secref{subs:qfi_indirect} we 
consider indirect estimation of the separation via estimation of the purity, relative intensity and coherence factor 
of the corresponding density operator. \secref{sec:conclusions} contains a summary of our findings and concluding remarks.

\section{\label{sec:Background} Background}

In this section we provide a rigorous mathematical background of the problem of imaging partially coherent point 
sources both in the classical (\secref{sec:classical_imaging}) as well as in the quantum mechanical case 
(\secref{sec:quantum_imaging}).  In \secref{sec:qubit_model} we introduce the qubit 
model of \cite{Chrostowski2017}, which provides a simplified description of the imaging problem, whose validity 
we will elucidate on in \secref{sec:qubit_model_validity}. Finally, \secref{subs:param_est} reviews the key 
concepts in both classical and quantum statistical inference.
 
\subsection{\label{sec:classical_imaging}Classical description of imaging of point sources}

Imaging systems have played a foundational role in the development of science since their invention in the late 
sixteenth century. Such systems typically consist of an aperture and a series of lenses designed to collect and 
analyze light emitted or reflected by an object, allowing for the study of the object's structure and properties. 
We model image formation in a simplified setting that neglects the imaging optics and sensor response, focusing
instead on the far-field diffraction pattern produced by the aperture. Throughout, we assume that the object 
consists of \emph{quasi-monochromatic point sources}, so that the image formed by each source is 
characterized by the system’s \emph{Point Spread Function} (PSF). A heuristic bound on the angular resolution 
imposed by diffraction was first introduced by Lord Rayleigh, who stated that two point sources can barely be 
resolved if the central maximum of one source’s PSF coincides with the first minimum of the other
in the image plane~\cite{Rayleigh80}.

This resolution criterion can be made quantitative using classical wave optics. For a circular aperture, the 
far-field diffraction pattern forms an Airy disk, whose first minimum occurs at an angle
	\begin{equation}
		\sin\phi\approx\phi \approx \frac{1.22 \lambda}{D} \,,
	\label{eq:diffraction_limit}
	\end{equation}
where \( \lambda \) is the wavelength of light, \( D \) is the diameter of the aperture, and \( 1.22 \) is the 
numerical factor arising from the first zero of the Bessel function, \( J_1 \), which characterizes diffraction 
from a circular aperture~\cite{Born1999}. More generally, under the Fraunhofer approximation, the complex field 
amplitude observed in the image plane is proportional to the Fourier transform of the aperture function. For a 
two-dimensional aperture described by a transmission function \( f(\bm{r}),\, \bm r\in\R^2 \), the far-field 
amplitude at the position \( \bm{r}' \in\R^2\) on the image plane is
	\begin{equation}
		\Psi(\bm{r}') \propto \int f(\bm{r}) \, \exp\left( -\ii\frac{k}{L} \, \bm{r} \cdot \bm{r}' \right) \, 
        \d^2 \bm{r} \,,
	\label{eq:complex_ampl}
	\end{equation}
where \( k = 2\pi/\lambda \) is the wavenumber, \( L \) is the distance from the aperture to the image plane, and 
\(\bm r'= L\sin\theta(\cos\phi,\sin\phi)\) describes points in the far field at angle \(\theta\) from the optical 
axis. The intensity at \( \bm{r}' \) on the image plane is proportional to the squared modulus of the complex 
field amplitude,
	\begin{equation}
		I(\bm{r}') \propto \abs{\Psi(\bm{r}')}^2 \,,
	\label{eq:intensity_r}
	\end{equation}
with the proportionality constant given by \((\lambda L)^{-1}\). The total optical power incident on the image 
plane is given by 
	\begin{equation}
		I = \int I(\bm{r}')\, \d^2 \bm{r}' \,.
	\label{eq:intensity_total}
	\end{equation}

If more than one source is being imaged, one must account for correlations between the complex amplitudes of the 
fields. These are captured by the \emph{mutual coherence matrix}, \( \Gamma_{ij} \in\C\), whose diagonal elements 
represent the average intensities of the individual sources, while the off-diagonal elements describe the degree of 
coherence between the various sources. Note that \( \Gamma_{ij} \) encodes the statistical properties of the sources 
and thus does not depend on the position \( \bm{r}' \) on the image plane; all spatial variation in the intensity arises from 
the complex field amplitudes \( \Psi_i(\bm{r}') \), which describe diffraction and propagation effects. If \(m\) 
quasi-monochromatic sources are being imaged, the intensity at position \( \bm r' \) on the image plane is given, 
under stationary statistical assumptions~\cite{Born1999}, by
	\begin{equation}
		I(\bm{r}') = \sum_{i,j=1}^m \Gamma_{ij} \, \Psi_i(\bm{r}') \, \Psi_j^*(\bm{r}') \, .
	\label{eq:partial_coherence_intensity}
	\end{equation}
Defining the \emph{complex degree of coherence}, $\gamma_{ij} = \frac{\Gamma_{ij}}{\sqrt{\Gamma_{ii} \Gamma_{jj}}},\, |\gamma_{ij}| \leq 1$,
\eqnref{eq:partial_coherence_intensity} can be rewritten as
	\begin{equation}
		I(\bm{r}') = \sum_{i,j=1}^m  \sqrt{\Gamma_{ii} \Gamma_{jj}} \, \gamma_{ij} \, \Psi_i(\bm{r}') \, \Psi_j^*
        	(\bm{r}') \, ,
	\label{eq:intensity_normalized_coherence_corrected}
	\end{equation}
with the total intensity given by \eqnref{eq:intensity_total}.

\subsection{\label{sec:quantum_imaging}Quantum mechanical treatment of imaging}

A quantum mechanical description of the imaging set-up described above proceeds by first associating the complex 
amplitude in \eqnref{eq:complex_ampl} to the position-space wavefunction of the photons after passing through the 
imaging system.  Defining the creation operator 
	\begin{equation}
		a^\dagger_{\Psi(\bm{r}')} = \int\, \d^2 \bm r' \Psi(\bm r') a^\dagger (\bm r') 
	\label{eq:second_quantization} 
	\end{equation}
with respect to the creation and annihilation operators at the image plane obeying $[a(\bm r), a^\dagger (\bm r')]=\delta^{(2)}(\bm r - \bm r')$,
the Fock state with exactly $n$ bosons in mode $\Psi(\bm r')$ at the image plane is given by 
	\begin{equation}
    \begin{split}
		\ket{\psi_n} &= \frac{a^{\dagger\, n}_{\Psi(\bm{r}')}}{\sqrt{n!}}\, \ket{0} \\
        &=\int\, \left(\prod_{j=1}^n\d^2 \bm r'_j \Psi(\bm r'_j) \right)\ket{\bm r'_1,\ldots 
		\bm r'_n}_{\mathrm{Symm}} \, ,
    \end{split}
	\label{eq:n_photon_state}	 
	\end{equation}
where $\ket{0}$ denotes the vacuum state and 
	\begin{equation}
		\ket{\bm r'_1,\ldots \bm r'_n}_{\mathrm{Symm}} = \frac{1}{\sqrt{n!}}\sum_{\pi\in\mathrm{S}_n}
		\ket{\bm r'_{\pi(1)}}\otimes\ldots\otimes\ket{\bm r'_{\pi(n)}}
	\label{eq:Fock_space_n}
	\end{equation}	
is a totally symmetric state. The position eigenkets at the image plane are defined as $a^\dagger(\bm r)\ket{0} = \ket{\bm r} $. 
The second equality in \eqnref{eq:n_photon_state} gives the position-space representation of the corresponding symmetric wavefunction 
\(\Psi(\bm r'_1\ldots\bm r'_n)\) and the state is normalized, \(\inner{\psi_n}{\psi_n}=1\).  For a chaotic quasi-monochromatic source the 
quantum state describing the field at the image plane is given by the density matrix
	\begin{equation}
		\xi = \sum_{n=0}^\infty \, p_n \ketbra{\psi_n}{\psi_n}
	\label{eq:quasi-mono_state}	
	\end{equation}
and the average number of bosons per coherence time illuminating the image plane is 
	\begin{equation}
		\bar{n}= \sum_{n=1}^\infty n\, p_n\, .
	\label{eq:average_photons_one_source}
	\end{equation}
For a thermal source \(p_n = \frac{\bar{n}^n}{(1+\bar{n})^{n+1}}\), whereas for a Poissonian source 
\(p_n = e^{-\bar{n}}\frac{\bar{n}^n}{n!}\).  The latter describes a coherent source of light in the absence of an adequate phase reference.

For $m$ partially coherent, chaotic, and quasi-monochromatic sources the corresponding density matrix at the image plane is 
given by
	\begin{equation}
		\xi = \sum_{\bm{n}, \bm{n}'} \rho_{\bm{n}\bm{n}'}\ketbra{\psi_{\bm{n}}}{\psi_{\bm{n}'}}\, ,
	\label{eq:quasi-mono_m}
	\end{equation}	
where $\bm n = (n_1,\ldots, n_m)$ and \(\ket{\psi_{\bm n}} = \bigotimes_{k=1}^m \ket{\psi_{n_k}}\) with 
$\ket{\psi_{n_k}}$ a Fock state with exactly \(n_k\) bosons in spatial mode \(\Psi_k(\bm r')\) as in 
\eqnref{eq:n_photon_state}. Observe that the fields at the image plane need not be orthogonal, i.e.,
	\begin{equation}
		[a_{\Psi_i}, a_{\Psi_j}^\dagger] = \int \d^2\bm r'\, \Psi_i^*(\bm r') \Psi_j(\bm r') = c_{ij}\in \C \, ,
	\label{eq:non_orthogonality}	
	\end{equation}
implying that the creation and annihilation operators associated with these modes obey the canonical commutation 
relations only when the mode functions are orthonormal.

The state \(\xi\) in \eqnref{eq:quasi-mono_m} can also be expressed in terms of the \emph{total number} 
basis as  
	\begin{equation}
    		\xi = \sum_{n=0}^\infty p_n \rho^{(n)},
	\label{eq:total_photon_number_decomp}
	\end{equation}
where $n=\sum_{k=1}^m n_k$, and \(0 \leq p_n \leq 1\) is the probability of detecting exactly \(n\) bosons in total 
(\(\sum_n p_n = 1\)). Each \(\rho^{(n)}\) in \eqnref{eq:total_photon_number_decomp} is a normalized density 
matrix supported on the \(n\)-boson subspace and can be expanded as  
	\begin{equation}
    		\rho^{(n)} = \sum_{\substack{\bm n, \bm n' \\ |\bm n| = |\bm n'| = n}} \rho^{(n)}_{\bm n\bm n'} 
            \ketbra{\psi_{\bm n}}
		{\psi_{\bm n'}},
	\label{eq:density_matrix_n_photons}
	\end{equation}
where \(|\bm n| = \sum_{k=1}^m n_k\) is the total boson number. 

Just as in the classical case, the presence of multiple sources gives rise to correlations between the complex 
amplitudes of their corresponding quantum mechanical states at the image plane.  These correlations are captured 
by the image plane \emph{coherence matrix}, a Hermitian matrix \(\Gamma' \in \mathbb{C}^{m \times m}\), defined 
as   
	\begin{equation}
    		\Gamma'_{ij} := \tr(\xi\, a_{\Psi_i}^\dagger a_{\Psi_j})\, .
	\label{eq:coherence_matrix}
	\end{equation}
The diagonal elements \(\Gamma'_{ii}\) represent the mean photon number in mode \(\Psi_i\), whereas the 
off-diagonal terms \(\Gamma'_{ij}\) capture the coherence between different modes. The \emph{average total 
number} at the image plane is the trace of the coherence matrix:
	\begin{equation}
    		\bar{n} = \sum_{i=1}^m \Gamma'_{ii} = \tr(\Gamma') = \sum_{n=0}^\infty\, n p_n\,.
	\label{eq:total_photon_number_xi}
	\end{equation}
In the low-intensity limit, \(\bar{n}\ll 1\), where the probability of having more than one photon per coherence 
time interval becomes negligible, the state in \eqnref{eq:density_matrix_n_photons} can be approximated as 
	\begin{equation}
		\begin{split}
			\xi &= (1-\bar{n})\ketbra{0}{0} + \bar{n}\rho^{(1)} + \cO(\bar{n}^2)\\
			     &= (1-\bar{n})\ketbra{0}{0} + \bar{n} \sum_{i,j=1}^m \gamma_{ij} \ketbra{\psi_i}{\psi_j} + 
                 \cO(\bar{n}^2)\, ,
		\end{split}
	\label{eq:low_intensity_xi}
	\end{equation}
where \(\Gamma'_{ij}=\bar{n}\gamma_{ij}\) and the sum in the second line of \eqnref{eq:low_intensity_xi} is over 
all $m$-dimensional vectors $\bm n$ that have a single non-zero entry at position $n_i$, i.e., 
\(\ket{\psi_i}=a^\dagger_{\Psi_i(\bm r')}\ket{0}, \, i\in(1,\ldots m)\). Physically, this corresponds to a regime 
in which individual photon detections carry nearly all the available information, and quantum interference 
effects between distinct modes \(\Psi_i(\bm r')\) are directly encoded in the off-diagonal elements 
\( \gamma_{ij} \).  

The remainder of this work concerns the low intensity limit.  In addition, we shall consider a one-dimensional 
imaging system illuminated by light from two partially coherent, quasi-monochromatic point sources with Gaussian 
PSFs
	\begin{equation}
		\begin{split}
			\Psi_1(x):=\Psi(x) = \left(\frac{1}{2\pi\sigma^2}\right)^{1/4} e^{- \frac{(x_1-x)^2}{4\sigma^2}}\\
			\Psi_2(x):=\Phi(x) = \left(\frac{1}{2\pi\sigma^2}\right)^{1/4} e^{- \frac{(x_2-x)^2}{4\sigma^2}}\, ,
		\end{split}
	\label{eq:Gaussian_PSFs}	
	\end{equation}
with mean \( x_1\neq x_2\) and variance $\sigma^2$.  This case is particularly relevant to the problem of imaging 
a pair of partially coherent, low-intensity sources emitting quasi-monochromatic 
light~\cite{Larson2018,Tsang2019b,Hradil2019,Hradil2021,Liang2021,Karuseichyk2022}. The 
effective density matrix describing the optical field at the image plane in the regime where at most one boson 
arrives per coherence time interval, is given by 
	\begin{equation}
		\xi = (1-\bar{n})\ketbra{0}{0} + \bar{n} \rho^{(1)} ,
	\label{eq:xi_2_gaussians}
	\end{equation}
where \(\rho^{(1)}\) is the \emph{normalized} single-photon density matrix describing the conditional state of 
the optical field given that a single photon arrives at the detector. It describes a statistical mixture of 
emissions from the two sources and is given by 	
	\begin{equation}
		\begin{split}
			\rho^{(1)} &= \frac{1}{1 + 2 c\gamma_R\sqrt{q(1-q)}} \left(q \ketbra{\psi}{\psi} + (1-q)\ketbra{\phi}
            {\phi} \right.\\
			&\left.+ \sqrt{q(1-q)}\left(\gamma \ketbra{\psi}{\phi}+\text{h. c.}\right)\right)\, ,
		\end{split}
	\label{eq:rho_1}
	\end{equation}
with \(q, (1-q)\) the relative intensities of the sources \( \Psi, \, \Phi\) respectively, $\ket{\psi},\ket{\phi}$ their 
respective one-photon states, and $\gamma\in\C, 0\leq \abs{\gamma}\leq 1, \gamma=\gamma_R + \ii \gamma_I$ 
their \emph{coherence factor}. Note that, with respect to \eqnref{eq:low_intensity_xi},
$\gamma_{11}=q,\gamma_{22}=(1-q),\gamma_{12}=\gamma_{21}^*=\gamma$. The average intensity per coherence time 
\(\bar{n}\) is given~\cite{Tsang2019b} 
	\begin{equation}
		\bar{n} = \delta \left(1 + 2 c\gamma_R\sqrt{q(1-q)}\right),
	\label{eq:average_n_2_sources}
	\end{equation}
where  
	\begin{equation}
		c = \abs{\int_{-\infty}^\infty \d x\, \Psi^*(x)\Phi(x)}=e^{-\frac{s^2}{8\sigma^2}}, \, s:=\abs{x_1-x_2}
	\label{eq:overlap}
	\end{equation}
is the overlap between the two PSFs and quantifies their spatial indistinguishability.  The 
low-intensity limit corresponds to \(\delta\ll1\), and physically corresponds to the situation where the  
average number of photons per coherence time interval arriving at the image plane due to only one source ($q=0,1$) is $\delta$, 
which we will assume known.
It is important to note that the intensity per coherence time interval fluctuates, and this fluctuation carries 
information about the separation \( s\), relative intensity \(q\), as well as coherence factor \(\gamma\).

\subsection{\label{sec:qubit_model}The qubit model~\cite{Chrostowski2017}}

The density matrix \( \rho^{(1)} \) of \eqnref{eq:rho_1} has support only on a two-dimensional subspace of the 
single-photon Fock space spanned by the PSFs. Denoting this subspace by \( \cH^{(1)} = \mathrm{span}
\{\ket{\psi},\,\ket{\phi}\} \), and introducing an orthonormal basis 
\(\{\ket{v_1}, \ket{v_2}\} \in \cH^{(1)} \), the state \( \rho^{(1)} \) can be expressed in Bloch form as
	\begin{equation} 
    		\rho^{(1)} = \sum_{i,j=1}^2 \rho_{ij} \ketbra{v_i}{v_j} = \frac{\one + \mathbf{r} \cdot \bm{\sigma}}{2},
	\label{eq:rho_qubit}
	\end{equation}
where \( \rho_{ij} = \braket{v_i|\rho^{(1)}|v_j} \), and \( \bm{\sigma} = (\sigma_1, \sigma_2, \sigma_3)^T \) is 
the vector of Pauli matrices defined with respect to the orthonormal basis \( \{\ket{v_1}, \ket{v_2}\} \). The 
vector \( \mathbf{r} \in \mathbb{R}^3 \)---known as the \emph{Bloch vector} of \( \rho^{(1)} \)---has components 
\( r_i = \tr(\sigma_i \rho^{(1)}) \) that quantify the contributions from source imbalance, mutual coherence, and PSF 
overlap. The magnitude of the Bloch vector satisfies \( 0 \leq r := |\bm{r}| \leq 1 \) and determines the purity 
of \( \rho^{(1)} \) via \( r^2 = 2\tr(\rho^{(1)2}) - 1 \).

The Bloch representation of \(\rho^{(1)}\) depends on the choice of orthonormal basis 
\( \{\ket{v_1}, \ket{v_2}\} \in \cH^{(1)} \). A particularly convenient choice of orthonormal basis is 
	\begin{equation}
		\begin{split} 		
			\ket{v_1^{(\alpha)}} &= \frac{1}{\sqrt{1 -2\alpha(1-\alpha)(1-c)}} \left(\alpha\ket{\psi} + (1-\alpha)\ket{\phi}\right) \\
    			\ket{v_2^{(\alpha)}} &= \frac{(1-\alpha(1-c)) \ket{\psi} -(c+\alpha(1-c)) \ket{\phi}}{\sqrt{(1-c^2)(1-2\alpha(1-\alpha)(1-c))}}.
	\end{split}
	\label{eq:centroid_basis}
	\end{equation}
with $0 \leq \alpha \leq 1$. In the limit \(s/\sigma\ll1\) the state \(\ket{v_1}\) in \eqnref{eq:centroid_basis} is equal to the zeroth order 
Hermite-Gauss mode, \(\ket{\psi(x_0)}=\int \d x \,\Psi_{x_0}(x) \ket{x}\) centred at the position \(x_0=\alpha x_1 +(1-\alpha) x_2\) 
between the two sources (see \eqnref{eq:Gaussian_PSFs}). Equivalently, \(\inner{v_1}{\psi(x_0)}=1+\cO(s^3)\). A second choice of orthonormal 
basis is given by 
	\begin{equation}
		\begin{split}
			\ket{e_1}&=\frac{1}{\sqrt{2(1-c)}}\left(\ket{\psi}-\ket{\phi}\right)\\
			\ket{e_2}&=\frac{1}{\sqrt{2(1+c)}}\left(\ket{\psi}+\ket{\phi}\right) 
		\end{split}
	\label{eq:geometric_basis}
	\end{equation}	
which can be obtained in the regime $s/\sigma\ll 1$ from \eqnref{eq:centroid_basis} by setting \(\alpha=1/2\).  In the limit \(s/\sigma\ll1\) the 
state \(\ket{e_2}\) in \eqnref{eq:geometric_basis} is equal to the zeroth order Hermite-Gauss mode, 
\(\ket{\psi(x_g)}=\int \d x \,\Psi_{x_g}(x) \ket{x}\) centred at the \emph{geometric centroid} 
\(x_g= (x_1 + x_2)/2\), i.e., \(\inner{e_2}{\psi(x_g)}=1+\cO(s^3)\). With respect to this basis the Bloch vector 
\(\rho^{(1)}\) is given by 
	\begin{equation} 
   		\mathbf{r} = \frac{-1}{1 + 2c\gamma_R \sqrt{q(1-q)}}
   		\begin{pmatrix}
       		(1-2q)\sqrt{1-c^2} \\
       		2\gamma_I \sqrt{q(1-q)} \sqrt{1-c^2} \\
       		c + 2\gamma_R \sqrt{q(1-q)}
   		\end{pmatrix},
	\label{eq:Bloch_vector}
	\end{equation}
with norm
	\begin{equation}
		r^2 = \frac{r^2_{\mathrm{inc}} + 4q(1-q)(\gamma_R^2+(1-c^2)\gamma_I^2)+4c\gamma_R\sqrt{q(1-q)}}
        {(1+2c\gamma_R\sqrt{q(1-q)})^2}\, 
	\label{eq:r}
	\end{equation} 
where, 
	\begin{equation}
		r^2_{\mathrm{inc}} = 1-4q(1-q)(1-c^2)\, ,
	\label{eq:r_incoh}
	\end{equation}
is the norm of the Bloch vector describing the incoherent mixture (\(\gamma=0\)) of point sources. 

The quantum state at the image plane---for both the fully quantum and qubit models---depends on five parameters: 
separation \( s \), relative intensity \( q \), coherence \( \gamma = \gamma_R + i\gamma_I \), and a reference position, $x_0$, 
from which to locate the positions of each PSF (such as the intensity centroid $x_c$ or the geometric centroid $x_g$). 

\subsection{\label{subs:param_est}Classical and quantum statistical inference}

We are thus led to the problem of estimating the key parameters of the PSF of a pair of partially coherent point 
sources, which can be grouped into the parameter vector \( \bm{\theta} = (x_0, s, q, \gamma_R, \gamma_I)^T \in \Theta \), 
where \( \Theta \) denotes the differentiable \textit{parameter space}. The task of estimating \( \bm{\theta} \) falls within the scope of 
classical and quantum parameter estimation theory, which we now briefly review.

\subsubsection{Classical parameter estimation}

Consider \( n \) independent and identically distributed (i.i.d.) samples \( \bm x = (x_1, \dots, x_n) \) drawn 
from a random variable \( X \sim p(x|\bm\theta) \), where the unknown parameters \( \bm\theta \in \Theta \) are 
to be estimated using an \textit{estimator} \( f : X^n \to \Theta \), which maps the data to an estimate 
\( \hat{\bm\theta} = f(\bm x) \). Any good estimator must satisfy the following four conditions~\cite{Wilks1962}:
	\begin{enumerate}[label=\roman*.]
    		\item \textit{Unbiasedness:} the estimator must give the correct parameters on average,
    			\begin{equation}
        				\braket{\hat{\bm\theta}} \coloneqq \int_{X^n} \text{d}^n\bm x\, p(\bm x|\bm\theta)
                        f(\bm x) = \bm\theta .
    			\label{eq:unbiasedness}
			\end{equation}
    		\item \textit{Consistency:}  for any \(\delta>0\) and a sequence \(f^{(k)}(\bm x)\) of estimates it 
            holds 
			\begin{equation}
				\lim_{k\to\infty}\, \mathrm{Pr}\left(\abs{\bm f^{(k)}(\bm x)-\bm\theta}>\delta\right)=0 .
			\label{eq:consistent}
			\end{equation}
    		\item \textit{Precision:} the estimator must minimize the error on its estimate as much as possible. 
            This error is quantified by the \textit{covariance matrix}, whose entries are
    			\begin{equation}
        				\hspace{7mm}\text{Cov}(f)_{jk}\coloneqq \int_{X^n} \text{d}^n\mathbf{x}(f_j(\mathbf{x})-
                        \theta_j)(f_k(\mathbf{x})-\theta_k)\, ,
    			\label{eq:precision}
			\end{equation}
    		where \(\theta_j \in \bm\theta\) is the \(j^\mathrm{th}\) parameter and 
            \(f_k(\bm x)\coloneqq\hat{\theta}_k\in\hat{\bm\theta}\) the \(k^{\mathrm{th}}\) element of our 
		      estimate.
    		\item \textit{Efficiency:}  The estimator \(f:X^n\rightarrow\Theta\) is said to be efficient if for 
            any other estimator \(g:X^n\to\Theta\) it holds 
			\begin{equation}
				\mathrm{Cov}(f)\mathrm{Cov}^{-1}(g)<\one\, .
			\label{eq:efficient}
			\end{equation}  
    \end{enumerate}
Here and throughout matrix inequalities are to be understood as follows: \(A \geq B \Leftrightarrow A-B\) is a 
positive semi-definite matrix.

A lower bound to the covariance matrix of any  unbiased estimator was provided by Cram\'er and
Rao~\cite{Cramer1961},
	\begin{equation}
    		\text{Cov}(f) \geq \bm F^{-1}[p(\bm x|\bm\theta)]\, ,
	\label{eq:CR_bound}
	\end{equation}
where \(\bm F[p(\bm x|\bm\theta)]\) is the \textit{Fisher Information matrix} (FI for short)~\cite{Fisher1922}
	\begin{equation}
        		\mathbf{F}_{jk}[p(\bm x|\bm\theta)]\coloneqq \int_{X^n} \text{d}\bm x p(\bm x|\bm\theta) 
                s_j(\bm x|\bm\theta) s_k(\bm x|\bm\theta)\, .
	\label{eq:Fisher_info}
	\end{equation}
The FI matrix is the covariance of the \emph{score} \(\bm s(\bm x|\bm\theta)\) of the random 
variable, a vector with elements \(s_j(\bm x|\bm\theta) = \frac{\partial \log p(\bm x|\bm\theta)}
{\partial\theta_j}\), and quantifies how much information the random variable $X$ carries about 
the parameters $\bm\theta$. Observe that for $n$ i.i.d. random variables, FI is additive i.e., 
\( \bm F[p(\bm x|\bm\theta)]=n \bm F[p(x|\bm\theta)]\).  The maximum likelihood estimator is asymptotically 
efficient, meaning that its covariance approaches the Cram\'er–Rao bound in the large-sample 
limit~\cite{Wilks1962}.

The above discussion holds when all pertinent parameters of the model need to be estimated. 
In many multi-parameter scenarios, however, only a subset of the parameters is of primary interest. 
The remaining parameters may either be known a priori or treated as nuisance parameters---quantities 
that affect the statistical model but are not themselves the target of inference. Denoting the vector of 
parameters by $\bm{\theta} = (\bm{\eta}, \bm{\zeta}) = (\eta_1,\ldots,\eta_k,\zeta_1,\ldots,\zeta_l)$,
where $\eta_i$ ($1 \leq i \leq k$) are the parameters of interest and $\zeta_j$ ($1 \leq j \leq l$) denote 
the remaining parameters of the model, the FI matrix for the full model can be written in block form as
	\begin{equation}
    		\bm F[p(x\vert\bm\theta)] = \left(\begin{matrix} \bm F_{\bm\eta\bm\eta}[p(x\vert\bm\theta)] & \bm F_{\bm\eta\bm\zeta}[p(x\vert\bm\theta)] \\
    		\bm F_{\bm\zeta\bm\eta}[p(x\vert\bm\theta)] & \bm F_{\bm\zeta\bm\zeta}[p(x\vert\bm\theta)] \end{matrix}\right)\, 
    	\label{eq:Fisher_nuisance}
	\end{equation}
where $\bm F_{\bm\eta\bm\eta}[p(x \vert \bm\theta)]$ is the $k \times k$ matrix with elements 
$F_{\eta_i,\eta_j}$ defined in Eq.~\eqref{eq:Fisher_info}, and analogously for the remaining blocks.

In the case where the parameters $\bm\zeta$ are known, \eqnref{eq:Fisher_nuisance} reduces to 
$\bm F[p(x\vert\bm\theta)]=\bm F_{\bm\eta\bm\eta}[p(x\vert\bm\theta)]$ and the Cram\'er-Rao 
bound takes the form given in \eqnref{eq:CR_bound}. In contrast, when the parameters $\zeta$ are 
nuisance parameters, the relevant block of the inverse Fisher information matrix in \eqnref{eq:Fisher_nuisance} 
is obtained via the Schur complement and reads
	\begin{multline}
    		[\bm F^{-1}[p(x\vert\bm\theta)]]_{\bm\eta\bm\eta} 
            =(\bm F_{\bm\eta\bm\eta}[p(x\vert\bm\theta)] \\ -
		\bm F_{\bm\eta\bm\zeta}[p(x\vert\bm\theta)]\bm F^{-1}_{\bm\zeta\bm\zeta}[p(x\vert\bm\theta)]\bm F_{\bm\zeta\bm\eta}
		[p(x\vert\bm\theta)])^{-1}\, ,
    	\label{eq:Fisher_nuisance_inv}
	\end{multline}
where the inverse is assumed to exist. Accordingly, the Cram\'er-Rao bound in the presence of nuisance parameters 
becomes $\text{Cov}(f) \geq[\bm F^{-1}[p(x\vert\bm\theta)]]_{\bm\eta\bm\eta}$.

\subsubsection{\label{sec:Quantum_inference}Quantum parameter estimation}

The transition from classical to quantum parameter estimation follows from the realization that the random 
variable \(X\) corresponds to the outcome of a quantum measurement---described by the resolution of the identity 
into a set of positive operators \(\{E_x >0\, \vert \sum_{x\in X} E_x=\one\}\)---whose probability 
distribution, \(p(x|\bm\theta)\), is given by Born's rule:
	\begin{equation}
    		p(x|\bm\theta) \coloneqq \tr[\rho(\bm\theta) E_x],
	\label{eq:Born_rule}
	\end{equation}
where \(\tr[\cdot]\) denotes the trace, and \(\rho(\bm\theta)\) is the density operator describing the state of 
some quantum system. Unlike the classical case, we now have an additional optimization to perform since as any 
resolution of the identity into a set of positive operators constitutes a valid quantum measurement.  
Specifically, we can search over the set of all possible quantum measurements (this set is compact) in order to 
find a measurement whose corresponding probability distribution yields the maximum FI. This allows us to define 
the \textit{Quantum Fisher Information} (QFI) matrix as the following maximization:
	\begin{equation}
    		\bm\cF[\rho(\bm\theta)] \coloneqq \max_{\{E_x>0\vert \sum_x E_x=\one\}}\bm F[p(x|\bm\theta)]\, .
	\label{eq:def_QFI}
	\end{equation}
The optimal measurement for estimating \(\bm\theta\) corresponds to the eigenprojectors of the operators 
\(\{\Lambda_{\theta_j} : \theta_j \in \bm\theta\}\), defined implicitly as
	\begin{equation}
    		\frac{\partial\rho(\bm\theta)}{\partial\theta_j} \coloneqq\frac{\rho(\bm\theta) \Lambda_{\theta_j} + 
            \Lambda_{\theta_j} \rho(\bm\theta)}{2} \, ,
	\label{eq:SLD}
	\end{equation}
known as the \textit{Symmetric Logarithmic Derivatives} (SLDs) of each parameter $\theta_j$~\cite{Braunstein1994}. 
The elements of the QFI matrix are obtained from the SLD operators by
    \begin{equation}
    		\boldsymbol{\cF}_{ij}[\rho(\bm\theta)] = \tr [\Lambda_{\theta_i} \rho(\bm\theta) \Lambda_{\theta_j}] .
	\label{eq:QFI_elements}
	\end{equation}
Due to this additional optimization we have the following chain of inequalities
	\begin{equation}
    		\text{Cov}(f) \geq \bm F^{-1}[p(x|\bm\theta)]\geq \bm{\cF}^{-1}[\rho(\bm\theta)]\, ,
	\label{eq:QCRB}
	\end{equation}
the latter of which is commonly known as the \textit{quantum Cramér-Rao bound} (QCRB)~\cite{Braunstein1994}. 
In particular, for the diagonal elements in \eqnref{eq:QCRB}:
 	\begin{equation}
     		\Delta\theta_j^2 \geq \mathbf{F}^{-1}_{jj}[p(x|\bm\theta)]\geq \boldsymbol{\mathcal{F}}^{-1}_{jj}
             [\rho(\bm\theta)]
 	\label{eq:QCRB_diagonal}
 	\end{equation}
where \(\Delta\theta_j^2\) is the \textit{variance} of the parameter \(\theta_j\).
The \textit{precision} in estimating parameter $\theta_j$, is defined as  
	\begin{equation}
    		H_j \coloneqq \frac{1}{\Delta\theta_j^2}\, .
    	\label{eq:precisions}
	\end{equation}
An analogous analysis holds in the quantum setting when some parameters are known or act as nuisance parameters, 
with the classical FI matrix replaced by the QFI matrix and the effective bound for the parameters of interest determined 
by the corresponding Schur complement.

In general, the SLD operators $\Lambda_{\theta_j}$ do \textit{not} commute, which implies that the QCRB cannot 
be attained \textit{simultaneously} for the respective parameters, not even asymptotically. A necessary and 
sufficient condition for the attainability of the QCRB is
	\begin{equation}
    		\tr \left[ \rho(\bm\theta) [\Lambda_{\theta_j}, \Lambda_{\theta_k}]  \right] = 0 ,
	\label{eq:QCRB_attainability}
	\end{equation}
which is less strict than the mere commutativity of the corresponding SLD operators~\cite{Ragy2016}.

If $\rho(\bm\theta)$ is a block-diagonal operator of the form 
	\begin{equation}
  		\rho(\bm\theta) = (1-\beta(\bm\theta)) \rho^{(0)}(\bm\theta)\oplus \beta(\bm\theta) \rho^{(1)}(\bm\theta)
  	\label{eq:rho_bd}
	\end{equation}
with $0 \leq \beta(\bm\theta) \leq 1$ and $\rho^{(0)}(\bm\theta),~\rho^{(1)}(\bm\theta)$ density operators over orthogonal 
subspaces, then the SLDs are also block diagonal:
	\begin{equation}
  		\Lambda_{\theta_i} = \Lambda_{\theta_i}^{(0)} \oplus
  		\Lambda_{\theta_i}^{(1)}\, .
  	\label{eq:SLD_bd}
	\end{equation}
From \cref{eq:SLD}, we have that
	\begin{equation}
  		\begin{split}
    			\Lambda_{\theta_i}^{(0)} &=\ell_{\theta_i}^{(0)} - \frac{1}{1-\beta(\bm\theta)}
    			\frac{\partial\beta(\bm\theta)}{\partial\theta_i} \mathds{1}\, , \\
    			\Lambda_{\theta_i}^{(1)} &=\ell_{\theta_i}^{(1)} + \frac{1}{\beta(\bm\theta)}
    			\frac{\partial\beta(\bm\theta)}{\partial\theta_i} \mathds{1}\, .
  		\end{split}
  	\label{eq:SLD_bd_value}
	\end{equation}
with $\ell_{\theta_i}^{(0)},~\ell_{\theta_i}^{(1)}$ the SLDs of $\rho^{(0)}(\bm\theta),~\rho^{(1)}(\bm\theta)$, respectively. 
After some algebra \eqnref{eq:QFI_elements} gives,
	\begin{equation}
  		{\bm\cF}[\rho(\bm\theta)] =(1-\beta(\bm\theta)) {\bm\cF}[\rho^{(0)}(\bm\theta)] + 
  		\beta(\bm\theta) {\cF}[\rho^{(1)}(\bm\theta)] +\mathbf{F}[\beta(\bm\theta)]\, ,
  	\label{eq:QFI_bd}
	\end{equation}
with ${\bm\cF}[\rho^{(0)}(\bm\theta)],~{\bm\cF}[\rho^{(1)}(\bm\theta)]$ the QFI matrices of 
$\rho^{(0)}(\bm\theta),~\rho^{(1)}(\bm\theta)$ respectively, defined element-wise as
	\begin{equation}
  		{\bm\cF}_{ij}[\rho^{(0)}(\bm\theta)] \coloneqq\tr \left[\ell_{\theta_i}^{(0)} \rho^{(0)}(\bm\theta)
    		\ell_{\theta_j}^{(0)}\right] \qquad \text{(c.f. $\rho^{(1)}(\bm\theta)$)}\, ,
  	\label{eq:QFI_bd_elements}
	\end{equation}
and $\mathbf{F}[\beta(\bm\theta)]$ the classical FI matrix of the binomial distribution $\{\beta(\bm\theta),~ 1-\beta(\bm\theta) \}$, 
whose elements read
	\begin{equation}
  		\mathbf{F}_{ij} \coloneqq\frac{1}{\beta(\bm\theta)(1-\beta(\bm\theta))}\frac{\partial\beta(\bm\theta)}{\partial\theta_i}
  		\frac{\partial\beta(\bm\theta)}{\partial\theta_j}\, .
  	\label{eq:FI_clas}
	\end{equation}
Applying \eqnref{eq:QFI_bd} to the density matrix in \eqnref{eq:xi_2_gaussians}, and noting that the vacuum state holds no information about the 
parameters, i.e., ${\bm\cF}[\proj{0}]=0$, the QFI matrix for the density operator $\xi$ describing the quantum state of two weak, 
partially coherent sources at the image plane of our imaging device is given by 
	\begin{equation}
    		{\bm\cF}[\xi(\bm\theta)]= \bar{n}(\bm\theta) {\cF}[\rho^{(1)}(\bm\theta)] +\mathbf{F}[\bar{n}(\bm\theta)]\, ,
	\label{eq:QFI_xi}
	\end{equation}
where we have identified the parameter dependent probability distribution $\beta(\bm\theta)$ with the probability 
$\bar{n}(\bm\theta)$ of having one photon arriving at the image plane and stress that this probability depends explicitly on the 
parameters $\bm\theta$.  We note that the form of \eqnref{eq:QFI_xi} was also derived in~\cite{Kurdzialek2022} for the case of 
balanced sources with a known degree of coherence.

\section{\label{sec:qfi} Theoretical limits to resolution}

In this section, we obtain the precisions, \eqnref{eq:precisions}, for the set of parameters 
$\bm \theta = (s,q,\gamma_R,\gamma_I)$, encoded in the density matrix $\xi(\bm \theta)$ of \eqnref{eq:xi_2_gaussians}.
We consider two approaches: direct estimation of the parameters (\secref{subs:qfi_direct}), and indirect estimation via the purity of 
$\rho^{(1)}(\bm\theta)$, from which parameters of interest—such as the source separation—are inferred using standard error propagation 
(\secref{subs:qfi_indirect}). Our analysis involves both the case of known and nuisance parameters. In \secref{sec:qubit_model_validity} we 
show when and under what conditions the qubit model of \secref{sec:qubit_model} can be safely employed in calculating the precision limits for 
the parameters $\bm{\theta}$.

\subsection{\label{subs:qfi_direct}Direct computation of precision}
Without loss of generality, we may choose the position
	\begin{equation}
		x_0 = \alpha x_1 + (1-\alpha)x_2 \coloneqq 0
	\label{eq:reference_position}
    \end{equation}
as the origin of coordinates in the image plane, with $\alpha \in [0,1]$.  With respect to this reference position 
the location of the sources are $x_1 = -(1-\alpha)s$ and $x_2 = \alpha s$ respectively. Note that choosing 
$\alpha = 1/2$ corresponds to setting the geometric centre, $x_g$, as the origin of coordinates, whereas selecting 
$\alpha = q$ corresponds to the intensity centroid, $x_c$. 

For any fixed choice of $\alpha \in [0,1]$ we note that the partial derivatives with respect to $q$, 
$\gamma_R$ and $\gamma_I$ remain in the span of the eigenvectors of $\rho^{(1)}$, i.e.,
	\begin{equation}
		\frac{\partial \rho^{(1)}}{\partial \theta_i} = \frac{1}{2}
		\frac{\partial \mathbf{r}}{\partial \theta_i} \cdot \boldsymbol{\sigma},\hspace{20pt} \theta_i 
		\in \{q, \gamma_R, \gamma_I\} \, ,
	\label{eq:dersqgamma}
	\end{equation}
with $\mathbf{r}$ defined in \eqnref{eq:Bloch_vector}. By decomposing the SLDs as
	\begin{equation}
		\Lambda_{\theta_i} = \lambda^{(0)}_{\theta_i} \openone +\boldsymbol{\lambda}_{\theta_i} \cdot \boldsymbol{\sigma} ,
		\hspace{20pt} \theta_i \in \{q,\gamma_R, \gamma_I\}\, ,
	\label{eq:SLD_qubit_1}
	\end{equation}
one arrives at the the following solutions
	\begin{equation}
		\begin{split}
			\lambda^{(0)}_{\theta_i}& =\frac{\frac{\partial\mathbf{r}}{\partial\theta_i} \cdot \mathbf{r}}{r^2 - 1}\\
			\boldsymbol{\lambda}_{\theta_i} & = \frac{\partial \mathbf{r}}{\partial \theta_i} - \lambda^{(0)}_{\theta_i} \mathbf{r}
		\end{split}\hspace{20pt}   \theta_i \in \{q, \gamma_R, \gamma_I\}.
	\label{eq:SLD_qubit_2}
	\end{equation}
The explicit dependence of \(\lambda^{(0)}_{\theta_i}\) and \(\boldsymbol{\lambda}_{\theta_i}\) on 
the parameters is provided in the \ref{app:SLD_computation}.

For the separation $s$, and reference position $x_0$ the determination of the relevant SLD is a bit more involved on account of the 
fact that the states \(\ket{\psi},\ket{\phi}\) appearing in \(\rho^{(1)}\) depend on both $s$ and $x_0$.  We will outline the procedure for 
obtaining $\Lambda_s$, with $\Lambda_{x_0}$ following similar reasoning (for the explicit derivation see the \ref{app:SLD_computation}). Working in 
the $\{\ket{e_1},\ket{e_2}\}$ basis of \eqnref{eq:geometric_basis} the derivative of \(\rho^{(1)}(\bm\theta)\)
with respect to $s$ is
    	\begin{equation}
		      \frac{\partial \rho^{(1)}}{\partial s} = \sum_{i,j=1}^2 \Big(\frac{\partial \rho_{ij}}{\partial s} 
            		\ket{e_i} \bra{e_j} +\rho_{ij} \ket{\partial_s e_i} \bra{e_j}+ \rho_{ij} \ket{e_i} \bra{\partial_s e_j}\Big)\,  , 
	\label{eq:deriv_rho_s}
    	\end{equation}
where we have introduced the shorthand notation
    \begin{equation}
	   \ket{\partial_s \psi} \coloneq \frac{\partial}{\partial s} \ket{\psi}= \int_{-\infty}^{+\infty} \text{d}x\, 
	   \frac{\partial \Psi(x)}{\partial s} \, \ket{x} \, .
	\label{eq:partial_der}
    \end{equation}
Note that $\{\ket{\partial_s e_1}, \ket{\partial_s e_2} \}\notin\cH^{(1)}$. Hence,
$\partial\rho^{(1)} / \partial s$ is an operator acting over a $4-$dimensional subspace spanned by
$\{\ket{e_1}, \ket{e_2}, \ket{\partial_s e_1}, \ket{\partial_s e_2}\}$. Using the Gram-Schmidt method, 
we can build an orthonormal basis $\{\ket{e_1}, \ket{e_2}, \ket{e_3}, \ket{e_4}\}$ of this extended subspace 
(preserving the original definitions of $\ket{e_1}$ and $\ket{e_2}$), relative to which $\Lambda_s$ can be 
expressed as
    \begin{equation}
	   \Lambda_s =\left(\begin{array}{c|c}
			         \Lambda_{11}^{\text{(qb)}}+\Lambda_{11}^{\text{(ex)}} & \Lambda_{12}\\
			         \hline \Lambda_{12}^{\dagger}& \Lambda_{22}\end{array}\right)\,,
	\label{eq:SLD_s}
    \end{equation}
where \(\Lambda_{11}^{\text{(qb)}},\Lambda_{11}^{\text{(ex)}}:\cH^{(1)}\to\cH^{(1)}\), 
\(\Lambda_{22}:\cH^{(2)}\to\cH^{(2)}\), and \(\Lambda_{12}:\cH^{(2)}\to\cH^{(1)}\), and we
have defined $\cH^{(2)}:=\mathrm{span}\{\ket{e_3},\,\ket{e_4}\}\). $\Lambda_s$ is to be understood as the sum of 
two contributions: a \textit{qubit term} ($\Lambda_{11}^{\text{(qb)}}$), corresponding to the SLD if 
\(\ket{\psi},\,\ket{\phi}\) did not depend on $s$, and the extra terms 
\(\Lambda_{11}^{\text{(ex)}},\,\Lambda_{12},\,\Lambda_{12}^{\dagger}$ and $\Lambda_{22}\), which exclusively 
contain the states' dependence on $s$.  $\Lambda_{x_0}$ can be similarly decomposed as in \eqnref{eq:SLD_s}. 
Once the SLDs have been found, the QFI matrix $\bm\cF[\rho^{(1)}(\bm\theta)]$ can be determined using \eqnref{eq:QFI_elements}, 
and the QFI matrix ${\bm\cF}[\xi(\bm\theta)]$ can be readily obtained through \cref{eq:QFI_xi}, since as
	\begin{equation}
    		\mathrm{F}_{\theta_j \theta_k}[\bar{n}(\boldsymbol{\theta})] =\frac{1}{\bar{n}(\boldsymbol{\theta}) \left(1-\bar{n}(\boldsymbol{\theta}) \right)}
		\frac{\partial \bar{n}(\boldsymbol{\theta})}{\partial\theta_j}\frac{\partial \bar{n}(\boldsymbol{\theta})}{\partial \theta_k}
	\label{eq:QFI_binomial}
	\end{equation}
is the classical FI matrix of the binomial distribution $\{\bar{n}(\bm\theta),~1-\bar{n}(\bm\theta)\}$.

\begin{figure}[ht!]
    \centering
    \includegraphics[width=\linewidth]{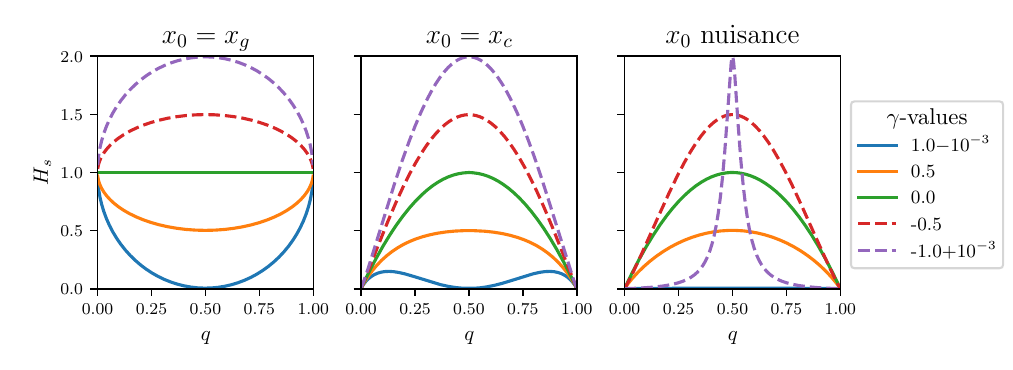}
    \caption{Precision for estimating the separation when $x_0$ is the geometric center of the sources (left), when it is the intensity centroid (middle), and when it is an unknown nuisance parameter (right). The precisions are plotted as functions of $q$ and for a finite separation $s/\sigma = 0.1$, treating the parameters $q$, $\gamma_R$ and $\gamma_I$ as known. The plots are done for purely real values of $\gamma$, as denoted by the legend---the left and middle plots are independent of $\gamma_I$, whereas the precision when $x_0$ is a nuisance parameter decreases with $|\gamma_I|$. $H_s$ is in units of $(\delta/4\sigma^2)$, and the numerical computation has been done for $\delta=10^{-2}$.}
    \label{fig:precisions_comp}
\end{figure}

Note that $\Lambda_s$ is reference-frame dependent, even at small separations. This dependence also affects the 
precision for the separation $H_s$, as shown in
\cref{fig:precisions_comp}. 
Assuming all other parameters known, and taking $x_0$ as the geometric center between the sources (i.e. $\alpha = 1/2$) results in $H_s$ being strictly finite 
even when only a single source is present, i.e., $q=0,1$.
This is due to the fact that, as $q\to0,1$, the separation can 
still be determined from the difference in position between the geometric centre and the brightest source. Choosing 
the intensity centroid, $\alpha=q$, as our reference position results in $H_s$ being more pronounced 
the more balanced, and anti-correlated the point sources are, whilst tending to zero when only 
a single source is present, i.e., $q=0,1$.  This is due to the fact that as $q\to0,1$, the intensity centroid tends to merge with the 
position of the brighter source, making separation estimation impossible. The only case where super-resolution is not possible is
the limiting case where $\gamma=1$ and $q=1/2$. \cref{fig:precisions_comp} also shows the precision $H_s$ for the case where $x_0$ is an unknown nuisance parameter. For the 
latter case, observe that super-resolution is still possible except in the limiting cases $q=\{0,1\}$ and $\gamma=\{1,\pm i\}$. Moreover, the lack of knowledge of $x_0$ hinders our super-resolving capabilities, as the precision $H_s$ is generally lower in this case. Remarkably, for incoherent sources, the precisions when $x_0$ is known to be the centroid and when it is a nuisance parameter coincide.

\begin{figure}[!htb]
    \begin{center}
    	\includegraphics[width=\columnwidth]{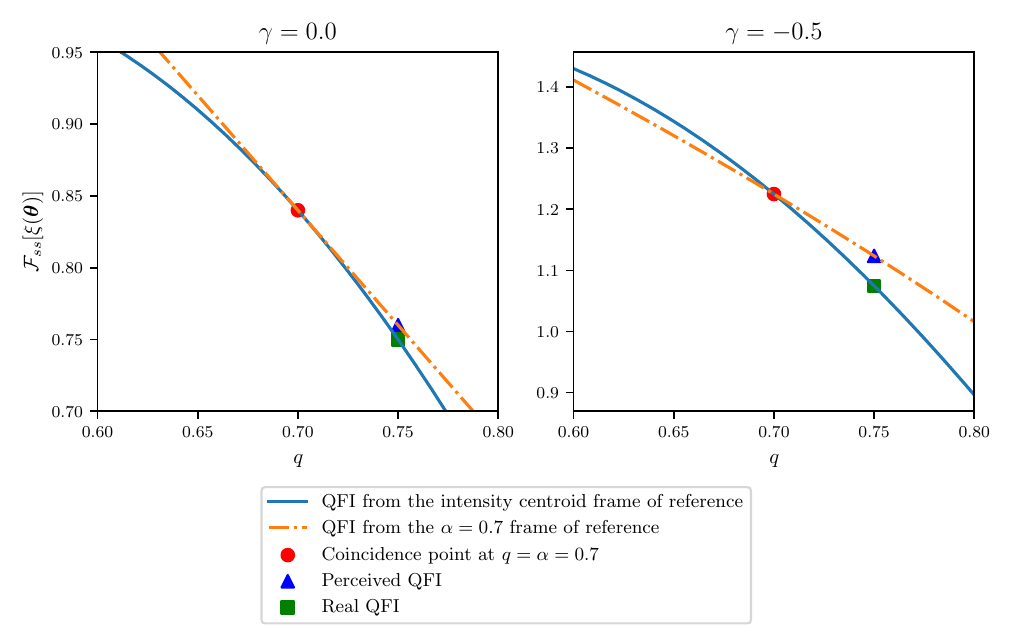}
    \end{center}
    \caption{Comparison between the true QFI for $s$ measured from the intensity centroid (solid blue line) and 
    the one measured from the $\alpha=0.7$ frame (orange dashed line). The QFI is in units of $(\delta/4\sigma^2)$, 
    with the computation being done for $\delta=10^{-2}$. For this particular example we consider $q=0.75$ the true value of the relative intensity, 
    and $\hat{q}=q+\epsilon(q)=\alpha=0.7$ the estimate for which we set the reference frame. The true QFI for $s$ 
    at $q=0.75$ is marked by the green square, whereas the perceived QFI from the $\alpha=\hat{q}=0.7$ reference 
    frame is marked by the blue triangle. For $\gamma = 0$ or $q = 1/2$ the difference between the real and perceived 
    QFIs is of order $\cO(\epsilon^2(q))$ (left plot), otherwise their difference is of order $\cO(\epsilon(q))$ (right plot).}
  \label{fig:QFI_alpha_vs_centroid}
\end{figure}

We pause here briefly to raise an important remark concerning the use of the intensity centroid as reference frame.  
Naively setting $\alpha=q$ in \eqnref{eq:reference_position} induces a spurious $q$-dependence over the modes of 
\eqnref{eq:Gaussian_PSFs}. This is unrealistic since as all spatial variations in intensity are strictly due 
to the mode wavefunctions, with $q$ governing the statistical properties of the sources (which are independent 
of the image plane position). To resolve this issue one notes that the intensity centroid is an inferred quantity, 
represented by an \emph{unbiased} estimator $\hat{q}$ with an associated error $\epsilon(q)$, 
i.e., $\alpha=\hat{q}=q \pm \epsilon(q)$. In \cref{fig:QFI_alpha_vs_centroid} we compare the QFI matrix element 
$\cF_{ss}[\xi(\bm\theta)]$ for both $\alpha=q$, corresponding to perfect alignment with the reference position, 
and $\alpha=\hat{q}$ corresponding to slight misalignment with respect to the reference position by 
$\pm\epsilon(q)$~\cite{deAlmeida21,Gessner20,Grace20}. We observe that the error propagated due to misalignment 
is of order $\cO(\epsilon(q))$ (at the limit $s/\sigma \ll 1$), unless $\gamma=0$ or $q=1/2$, where said error is of order 
$\cO(\epsilon^2(q))$. Henceforth, we shall set $\alpha=\hat{q}=q$, whenever we choose the intensity centroid as our reference position, 
carefully neglecting any spurious $q$-dependence on the modes.

\begin{figure}[ht!]
	\begin{center}
    		\includegraphics[width=\columnwidth]{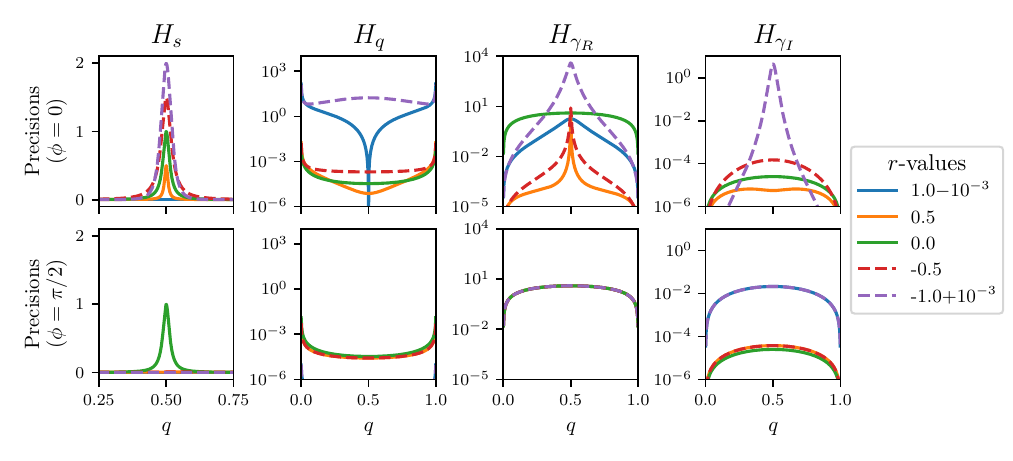}
  	\end{center}
  	\caption{Precisions of the parameters ${\bm\theta} = (s,q,\gamma_R,\gamma_I)$ as a function of the relative intensity $q$ 
	for a separation $s/\sigma = 0.1$.  Al remaining model parameters are treated as nuisance parameters. $H_s$ is in units of 
	$(\delta/4\sigma^2)$, whereas all other precisions are dimensionless. The numerical computation has been done for $\delta=10^{-2}$.
	The plots are done for several values of $\gamma\coloneqq r e^{i\phi}$, with the top four plots corresponding 
  	to $\phi=0$ and the bottom four to $\phi=\pi/2$, and the values of $r$ as denoted by the legend.
	The precision $H_s$ is plotted for $1/4\leq q \leq 3/4$ to better highlight the central peak.}
  \label{fig:precisions_full}
\end{figure}

\cref{fig:precisions_full} displays the precisions, given by \cref{eq:precision}, for each of the four parameters 
$(s,q,\gamma_R,\gamma_I)$, as a function of the relative intensity $q$, in the sub-Rayleigh regime ($s/\sigma = 0.1$) . 
Here, we treat all other parameters as nuisance parameters. The precisions are computed for the case of purely real 
and purely imaginary values of the coherence factor $\gamma$ to explicitly illustrate the roles these two parameters play 
in estimation precision. Whilst lack of knowledge of $\gamma$ does affect resolution it does not cause a resurgence 
of Rayleigh's curse as claimed in~\cite{Larson2018, Liang2021}. For a small, but finite separation, $s/\sigma=0.1$ we find that 
$H_s$ is finite for values of $q$ close to $1/2$, and hence sub-diffraction imaging is possible. In fact, in the limit $s\to 0$ we observe that 
$H_s$ remains finite at $q=1/2$ and vanishes elsewhere. However, for $\gamma\to 1$ or $\gamma\to\pm i$, $H_s$ vanishes 
at small separations even for $q=1/2$. Hence, for these particular values of $\gamma$ super-resolution is not achievable. 

The parameter $\gamma_R$ can also be estimated in the $s\to 0$ limit, save for the cases $q=0,1$, as the precision 
$H_{\gamma_R}$ remains finite. On the other hand, the precisions $H_q$ and $H_{\gamma_I}$ vanish in the limit $s\to 0$ 
(with the exception of $H_q$ at the extreme cases $q=0,1$, where only one source is present). 
Finally, $\cF_{\gamma_I\gamma_I}[\xi(\bm\theta)]$ vanishes in the limit $s\to 0$, resulting in vanishing 
precision---the lack of knowledge of other parameters can only hinder our resolving capabilities. In \cref{fig:precisions_full} 
$H_q$ and $H_{\gamma_I}$ are shown finite due to us choosing $s/\sigma=0.1$.

\begin{figure}[ht!]
  \begin{center}
    \includegraphics[width=\columnwidth]{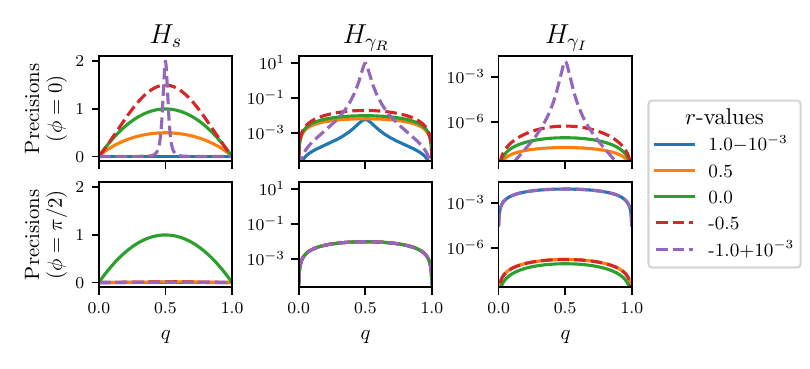}
  \end{center}
  \caption{Precisions of the parameters $s,\gamma_R,\gamma_I$, with $q$ assumed known. 
  $H_s$ is in units of $(\delta/4\sigma^2)$, whereas all other precisions are dimensionless. The numerical computation has been done for $\delta=10^{-2}$. All precisions are plotted as 
  functions of $q$ for a separation $s/\sigma = 0.1$, well below the Rayleigh limit. The plots are done for several values of $\gamma\coloneqq re^{i\phi}$, 
  with the three top subplots corresponding to $\phi=0$ and the three bottom ones corresponding to $\phi=\pi/2$, and the values of $r$ as denoted by the legend.}
  \label{fig:precisions_sri}
\end{figure}

If one or more of the parameters $q,~\gamma_R,~\gamma_I$ are known, then super-resolution for estimating the separation 
of the sources is possible for the entire range of values of the relative intensity $q$. In \cref{fig:precisions_sri} we plot the 
precisions $H_s,~H_{\gamma_R},~H_{\gamma_I}$ assuming that the relative intensity $q$ is known and under the same conditions 
as in \cref{fig:precisions_full}. We observe that super-resolution for separation occurs for all values of $q$, except for the 
limiting cases $\gamma\to 1$ and $\gamma\to\pm i$, where the precision vanishes for all $q$. Concerning the real part of the 
coherence, $H_{\gamma_R} > 0$ and it is independent of $\gamma_I$. On the other hand, $\gamma_I$ cannot be estimated: 
$H_{\gamma_I}$ vanishes in the limit $s\to 0$, as expected. As before, the reason why $H_{\gamma_I}$ appears finite is due to us 
choosing $s/\sigma=0.1$. 

The precisions $H_s$, $H_q$ and $H_{\gamma_R}$ are independent of $x_0$, whilst $H_{\gamma_I}$ depends on the choice 
of reference position, even when $x_0$ is a nuisance parameter. However, this dependence becomes irrelevant in the $s/\sigma \ll 1$ limit since as 
$H_{\gamma_I}\xrightarrow{s/\sigma\ll 1}0$. Assuming prior knowledge of $x_0$ results in higher attainable precisions, particularly for 
the separation which depends strongly on $x_0$.

\begin{figure}[htb!]
  \begin{center}
    \includegraphics[width=\columnwidth]{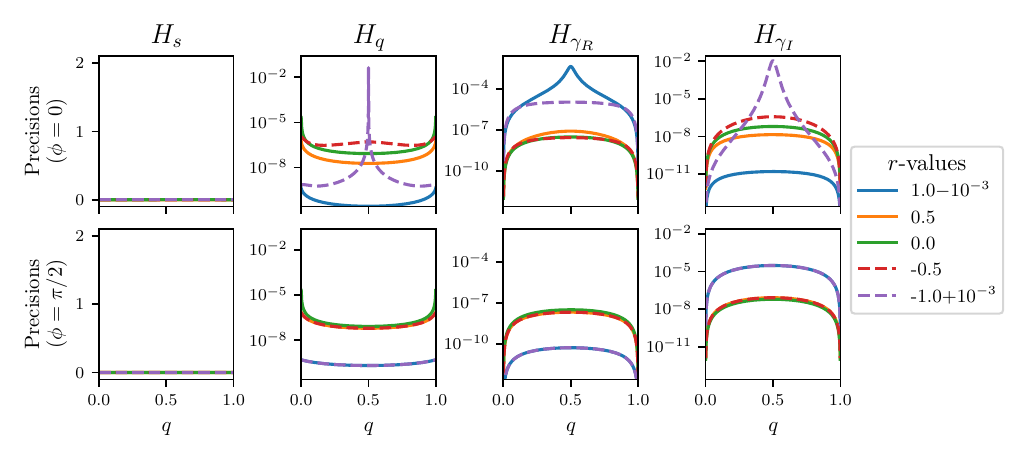}
  \end{center}
  \caption{Precisions of the parameters ${\bm\theta} = (s,q,\gamma_R,\gamma_I)$ as functions of $q$ for a separation $s/\sigma = 0.1$  Only the quantum term $\bar{n}(\bm\theta){\bm\cF}[\rho^{(1)}(\bm\theta)]$ 
   is considered for the QFI, thus resulting in a precision $H_s$ that vanishes in the limit $s\to 0$, for every $q$ and $\gamma$. 
   $H_s$ is in units of $(\delta/4\sigma^2)$, whereas all other precisions are dimensionless. The numerical computation has been done for $\delta=10^{-2}$. The plots are done for several 
  values of $\gamma\coloneqq  re^{i\phi}$, with the four top subplots corresponding to $\phi=0$ and the four bottom ones corresponding to $\phi=\pi/2$, and the 
  values of $r$ as denoted by the legend.}
  \label{fig:precisions_full_quantum}
\end{figure}

\begin{figure}[htb!]
  \begin{center}
    \includegraphics[width=\columnwidth]{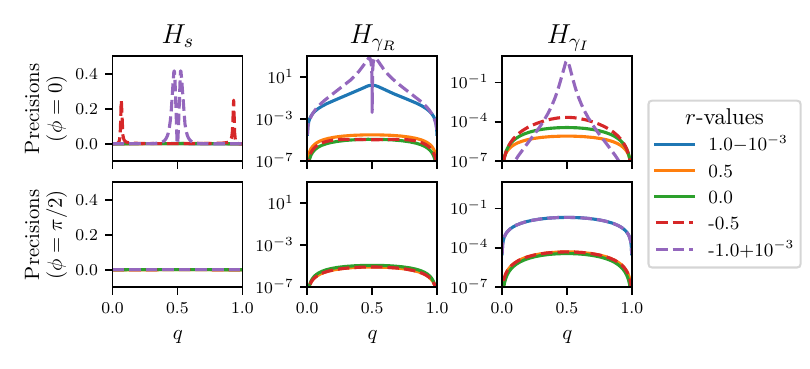}
  \end{center}
  \caption{Precisions of the parameters $s,\gamma_R,\gamma_I$, with $q$ assumed known and considering only the quantum term $\bar{n}(\bm\theta){\bm\cF}[\rho^{(1)}(\bm\theta)]$ for the QFI. $H_s$ is in units of $(\delta/4\sigma^2)$, whereas all other precisions are dimensionless. The numerical computation has been done for $\delta=10^{-2}$. All precisions are plotted as 
  functions of $q$ for a separation $s/\sigma = 0.1$, well below the Rayleigh limit. The plots are done for several values of $\gamma\coloneqq re^{i\phi}$, 
  with the three top subplots corresponding to $\phi=0$ and the three bottom ones corresponding to $\phi=\pi/2$, and the values of $r$ as denoted by the legend.}
  \label{fig:precisions_sri_quantum}
\end{figure}

Finally, in order to highlight the importance of the classical FI, $\mathbf{F}[\bar{n}(\bm\theta)]$, to the precisions of all pertinent 
parameters, we plot in \figsref{fig:precisions_full_quantum}{fig:precisions_sri_quantum} the precisions for all parameters of the model 
considering only the quantum contribution to the QFI, $\bar{n}(\bm\theta){\bm\cF}[\rho^{(1)}(\bm\theta)]$ (see \cref{eq:QFI_xi}), under the same conditions 
as in \figsref{fig:precisions_full}{fig:precisions_sri} respectively. Observe that in \figref{fig:precisions_full_quantum}---where we 
treat all remaining parameters as nuisance parameters---$H_s$ is zero for all values of the relative intensity $q$.  This is the result 
reported by Larson and Saleh~\cite{Larson2018} and Liang {\it et al.}~\cite{Liang2021}. We clearly see that the binomial distribution 
$\{\bar{n}(\bm\theta),1-\bar{n}(\bm\theta)\}$---arising from destructive interference due to the presence of coherence---contains crucial information that 
enhances and, in some cases restores, super-resolution. The same can be observed for the precision of $q$ and $\gamma_R$ with most of the information coming from the FI of the binomial distribution. \figref{fig:precisions_sri_quantum} demonstrates the same phenomenon for the case 
where $q$ is known.  In this case the single photon state does carry some information about the separation but not across the entire range of 
values for the relative intensity: Once again the information contained in the binomial distribution is crucial for the estimating the separation of 
the two sources. In particular, \figref{fig:precisions_sri_quantum} shows that $H_s$ vanishes in the limit $s\to0$ for equally bright sources.

\subsection{\label{sec:qubit_model_validity}Applicability and restrictions of the qubit model}
We now compute the precisions, \eqnref{eq:precision}, using the qubit model.  To do so we restrict ourselves 
entirely to the subspace $\cH^{(1)}$ so that the derivative of \(\rho^{(1)}(\bm\theta)\) with respect to any of 
the parameters is given by
    \begin{equation}
	   \frac{\partial \rho^{(1)}(\bm\theta)}{\partial\theta_i} = \frac{1}{2} \frac{\partial \mathbf r}
       {\partial\theta_i}\cdot\boldsymbol{\sigma}\, .
	\label{eq:derivative_qubit}
	\end{equation}
Whilst \eqnref{eq:derivative_qubit} is exact for the parameters \(q, \gamma_R, \gamma_I\), for the separation \(s\) 
and the reference position $x_0$ it ignores all contributions coming from the subspace spanned by the derivatives 
\(\{\ket{\partial_se_1}, \ket{\partial_se_2}\}\) and \(\{\ket{\partial_{x_0}e_1}, \ket{\partial_{x_0}e_2}\}\) 
respectively. Focusing exclusively on the separation, for the remainder of this section we shall 
assume that all remaining parameters are known so that the precision in estimating the separation of the sources is exactly 
equal to the QFI matrix element $\boldsymbol{\cF}_{ss}[\xi(\boldsymbol{\theta})]$. We define the qubit contribution to 
$\boldsymbol{\cF}_{ss}[\xi(\boldsymbol{\theta})]$ as
	\begin{align}\nonumber
		\boldsymbol{\cF}^{\text{(qb)}}_{ss}[\xi(\boldsymbol{\theta})]        
        & =\bar{n}(\boldsymbol{\theta})\boldsymbol{\cF}^{\text{(qb)}}_{ss}[\rho^{(1)}(\boldsymbol{\theta})]
		+ \bm F_{ss}[\bar{n}(\boldsymbol{\theta})] \\
		\boldsymbol{\cF}^{\text{(qb)}}_{ss}[\rho^{(1)}(\boldsymbol{\theta})] 
        & =\tr [{\Lambda_{11}^{\text{(qb)}}}^2 \rho^{(1)}(\boldsymbol{\theta})]\, ,
		\label{eq:qubit_contribution}
	\end{align}
with $\Lambda_{11}^{(\rm qb)}$ defined in \cref{eq:SLD_s} and the difference from the exact value of $\boldsymbol{\cF}_{ss}[\xi(\boldsymbol{\theta})]$ given by the contributions of the extra terms. 

\begin{figure}[htb!]
		\centering
		\includegraphics[width=\columnwidth]{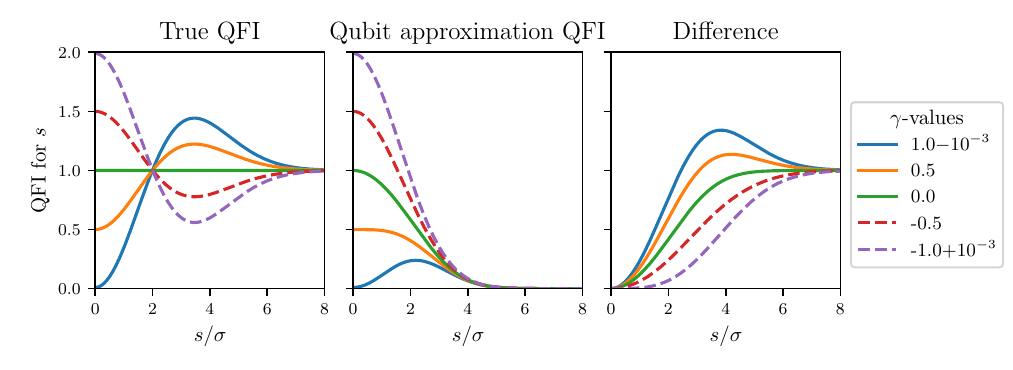}
		\caption{Qubit and extra contributions to the diagonal element corresponding to the separation for $\boldsymbol{\cF}_{ss}
        	[\xi(\boldsymbol{\theta})]$. The plots are done for $q=1/2$ and several real values of $\gamma$, as denoted by the legend. 
		The matrix element $\cF_{ss}[\xi(\bm\theta)]$ is independent of $\gamma_I$, hence our choice of purely real $\gamma$-values.
        	The value of $\boldsymbol{\cF}_{ss}[\xi(\boldsymbol{\theta})]$ is expressed in units of \((\delta/4\sigma^2)\) with the computations being done for $\delta=10^{-2}$, 
        	and the separation in units of $\sigma$.}
		\label{fig:QFI_diff_05}
	\end{figure}

Comparing the qubit contribution versus the total QFI element for the separation, we observe that the former can approximate 
the latter in the super-resolution regime $s/\sigma \ll 1$ (see \cref{fig:QFI_diff_05}). However, testing the validity of the 
\textit{qubit approximation}
	\begin{equation}
    		\cF_{ss}^{(\rm qb)}[\xi(\bm \theta)] \approx \cF_{ss}[\xi(\bm\theta)]\qquad \text{for} \qquad s/\sigma \ll 1
    \label{eq:qb_approx}
	\end{equation}
demands careful analysis, as the QFI matrix element for the separation is dependent on our choice of reference frame position $x_0$. 
Consequently, the approximation can hold in a given reference frame, but fail when working from a different one. For the qubit contribution  
given by the SLD $\Lambda_{11}^{(\rm qb)}$ in \cref{app:Lambda_11,app:Lambda_11explicit} of the \ref{app:SLD_computation}, we observe that the qubit approximation of \cref{eq:qb_approx} holds for any $q$ and $\gamma$, 
so long as the reference position $x_0$ is known to be the geometric centre of the sources. Note that for the computation 
of $\Lambda_{11}^{(\rm qb)}$ we have worked in the $\{\ket{e_1}, \ket{e_2} \}$ basis, which is centred precisely at the geometric centre of 
the sources (see the discussion below \cref{eq:geometric_basis}).
	
Similarly, we may work in the basis $\{\ket{v_1^{(\alpha)}}, \ket{v_2^{(\alpha)}} \}$, centred at the arbitrary position 
$x_0 = \alpha x_1 + (1-\alpha)x_2$ between the sources. This results in a different Bloch vector $\mathbf{r}$ and hence 
a different SLD $\Lambda_{11}^{(\rm qb)}$, such that the qubit term approximates the QFI element for $s$ corresponding to 
the same reference position $x_0$. For instance, working in the basis $\{\ket{v_1^{(\hat{q})}}, \ket{v_2^{(\hat{q})}}\}$ 
approximates, in the limit $s/\sigma \ll 1$, the QFI element for $s$ in  the intensity centroid reference frame, $x_0=x_c$.
In summary, the qubit approximation of \cref{eq:qb_approx} holds in any reference frame $x_0$, provided one works 
in the appropriate basis in the space of the modes. The first vector of such a basis must yield an average position equal to $x_0$ 
(the second vector is simply the orthogonal within the $2-$dimensional subspace). Any other choice of basis won't do.

To illustrate why the qubit approximation fails, consider the projective measurement associated to the projectors 
\(\{\ketbra{e_1}{e_1}, \ketbra{e_2}{e_2}\}\), and choose our reference position to be $x_0=x_c$.
As \(\ket{e_2}=\ket{\psi(x_g)}+\mathcal{O}(s^2)\) (the zeroth Hermite-Gauss mode centered at $x_g$)  the set of projectors \(\{\ketbra{e_1}{e_1}, \ketbra{e_2}{e_2}\}\) 
corresponds, in the low separation limit, to a binary SPADE measurement about the geometric centre of the two sources~\cite{Tsang16}. This measurement is 
known to be optimal when $q=1/2$ and $s/\sigma\ll1$.  Computing the partial derivatives with respect to the separation of the resulting 
probability distribution \(p(k\vert\bm\theta)=\tr(\ketbra{e_k}{e_k}\rho^{(1)}(\bm\theta))\) results in
	\begin{equation}
	   \begin{split}
		\frac{\partial p(k\vert\bm\theta)}{\partial s} & = \tr\left(\ketbra{e_k}{e_k}\frac{\partial\rho^{(1)}(\bm\theta)}{\partial s}\right)\\
		& = \frac{\partial\rho^{(1)}_{kk}(\bm\theta)}{\partial s} +(-1)^k\, 2\nu\Re(\rho^{(1)}_{12}(\bm\theta))
	\end{split}
	\label{eq:full_blown_derivative}
	\end{equation}
with \(\nu\) given in \cref{app:constants} of the \ref{app:SLD_computation}, and $\Re$ denoting the real part of a complex number. We note that no approximations 
have been made in deriving \eqnref{eq:full_blown_derivative}. If we now replace \(\frac{\partial\rho^{(1)}(\bm\theta)}{\partial s}\) with its qubit 
approximation \(\frac{1}{2}\frac{\partial\mathbf r}{\partial s}\cdot \boldsymbol{\sigma}\) we obtain instead
	\begin{equation}
		\begin{split}
			\frac{\partial p^{(\mathrm{qb})}(k\vert\bm\theta)}{\partial s} & = \tr\left(\ketbra{e_k}{e_k}\frac{1}{2}
			\frac{\partial\mathbf r}{\partial s}\cdot \boldsymbol{\sigma}\right) \\
		& = \frac{\partial\rho^{(1)}_{kk}(\bm\theta)}{\partial s}\, .
		\end{split}
	\label{eq:qubit_derivative}
	\end{equation}

\begin{figure}[htb!]
		\centering
		\includegraphics[width=\columnwidth]{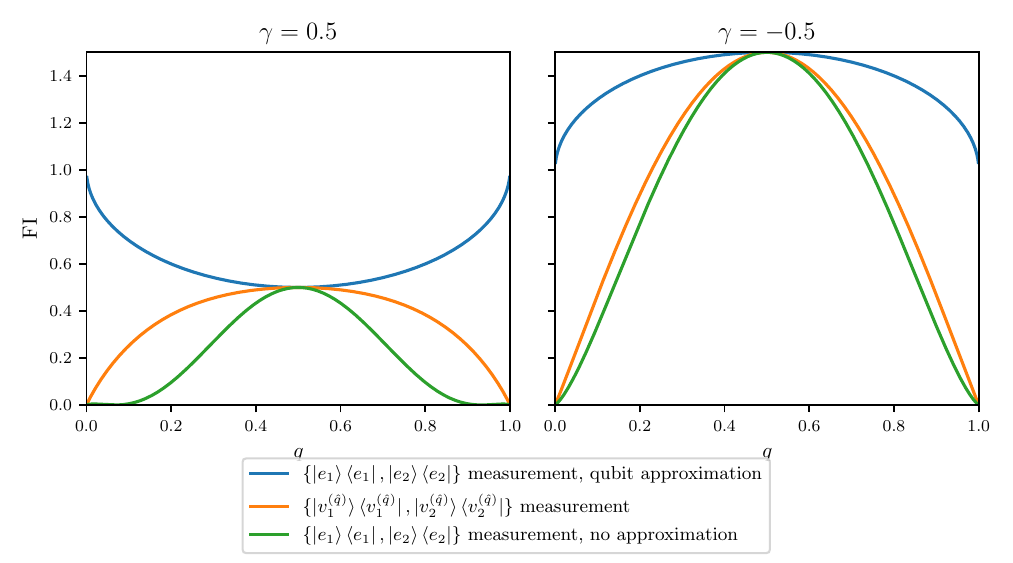}
		\caption{FI for the separation from the aligned SPADE measurement $\{\ketbra{v_1}{v_1},\ketbra{v_2}
        {v_2}\}$ and the misaligned one $\{\ketbra{e_1}{e_1},\ketbra{e_2}{e_2}\}$, with and without taking the 
        qubit approximation. The plots are made for $\gamma=0.5$ (left), $\gamma=-0.5$ (right), and
        $s/\sigma=10^{-2}$. The FI is in units of $(\delta/4\sigma^2)$, with $\delta=10^{-2}$. At small separations, 
        the FI of the aligned measurement coincides with the QFI and the FI of the qubit-approximated misaligned measurement 
        coincides with the failing qubit-approximated QFI of \cref{eq:qubit_contribution}.}
		\label{fig:misaligned_measurement_3}
	\end{figure}

The corresponding FI for the measurement \(\{\ketbra{e_1}{e_1}, \ketbra{e_2}{e_2}\}\) resulting from 
\eqnsref{eq:full_blown_derivative}{eq:qubit_derivative} is contrasted with the FI corresponding to the measurement 
\(\{\ket{v_1^{(\hat{q})}}\bra{v_1^{(\hat{q})}}, \ket{v_2^{(\hat{q})}}\bra{v_2^{(\hat{q})}}\}\) in \figref{fig:misaligned_measurement_3}. 
Note that the latter is equal to the QFI for this reference position and corresponds to a binary SPADE measurement 
centred at the intensity centroid $x_c$ when $s/\sigma \ll 1$. We observe that the qubit model erroneously over-estimates 
the FI, even beating the QFI. This is not surprising since as the FI of the \(\{\ketbra{e_1}{e_1}, \ketbra{e_2}{e_2}\}\) measurement 
matches (in the limit $s/\sigma \ll 1$) the QFI in the geometric centre reference frame, which is superior to the one from the intensity centroid frame (see \cref{fig:precisions_comp}).
	
The reason for this over estimation is of course the omission of the extra contributions due to \(\cH^{(2)}\), 
which are negative and reduce the FI.  For the case of the orthonormal basis \(\{\ket{v_1^{(\alpha)}}, \ket{v_2^{(\alpha)}}\}\) these 
contributions tend to zero in the limit $s/\sigma\ll 1$, since 
	\begin{equation}
	   \braket{v_j^{(\alpha)} | \partial_s v_k^{(\alpha)}}=\mathcal{O}(s^2), \, j\neq k\in\{1,2\} \,.
	\label{eq:qubit_model_valid}
	\end{equation}
It follows that the qubit model can be safely applied whenever, $\lim\limits_{s/\sigma\ll 1}\Pi_{\cH^{(1)}}\Pi_{\cH^{(2)}}=\mathcal{O}(s^2)$
with $\Pi_{\cH^{(1)}}$ (resp. $\Pi_{\cH^{(2)}}$) the projector onto the subspace $\cH^{(1)}$ (resp. $\cH^{(2)}$).
Note that, the $\{\ket{v_1^{(\alpha)}},\, \ket{v_2^{(\alpha)}}\}$ basis carries one degree of freedom, namely the position of the reference frame 
$x_0$, which, together with the three additional degrees of freedom of the qubit model, can encode a total of 
four parameters, one less than the full model. We will explore the consequences of this deficiency in \secref{subs:qfi_indirect}.

One can recover \eqnref{eq:qubit_derivative} if one performs the full derivative of both the measurement and 
$\rho^{(1)}$ without any approximation, i.e.,
	\begin{equation}
	   	\begin{split}
			\frac{\partial p(k\vert\bm\theta)}{\partial s} & = \tr\left(\frac{\partial}{\partial s}\left(\ketbra{e_k}{e_k}\right)
			\rho^{(1)}(\bm\theta)+\ketbra{e_k}{e_k}\frac{\partial\rho^{(1)}(\bm\theta)}{\partial s}\right) \\
			& =  \frac{\partial\rho^{(1)}_{kk}(\bm\theta)}{\partial s} 	\, .
		\end{split}
	\label{eq:full_blown_derivative2}
	\end{equation}
The reason for this is that the first term in \eqnref{eq:full_blown_derivative2} exactly cancels the last term in 
\eqnref{eq:full_blown_derivative}. In other words, the overestimation resulting from the qubit approximation is 
equivalent to that of performing a measurement that explicitly depends on the parameter,  which is 
unjustified as it implies previous knowledge on the parameter to be estimated.

\begin{figure}
		\centering
		\includegraphics[width=0.8\columnwidth]{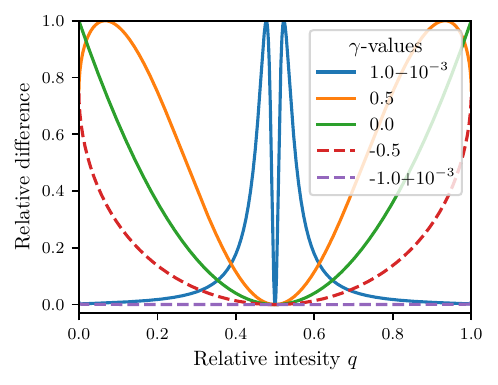}
		\caption{Relative difference between the Fisher information of the optimal measurement \(\{\ketbra{v_1}{v_1}, \, \ketbra{v_2}{v_2}\}\) 
		and the misaligned measurement \(\{\ketbra{e_1}{e_1},\, \ketbra{e_2}{e_2}\}\) as a function of the relative intensity. The plot is made 
		in the limit $s\to0$ and for several real values of $\gamma$, as denoted by the legend.}
		\label{fig:misaligned_measurement}
	\end{figure}

When \(q\neq 1/2\) a binary SPADE measurement centred anywhere but at the reference position $x_0$ (in our example $x_0=x_c$) can be thought of as a 
misalignment of the measurement apparatus, resulting in a sub-optimal FI. This is the reason why, 
in \cref{fig:misaligned_measurement_3} and when the qubit approximation is \emph{not} taken, the FI of 
the \(\{\ketbra{e_1}{e_1},\, \ketbra{e_2}{e_2}\}\) measurement (misaligned \& sub-optimal) is \emph{below} 
the FI of the \(\{\ket{v_1^{(\hat{q})}}\bra{v_1^{(\hat{q})}},\, \ket{v_2^{(\hat{q})}}\bra{v_2^{(\hat{q})}}\}\) measurement (aligned \& optimal). \figref{fig:misaligned_measurement} 
shows the relative difference in FI between said measurements.
The misaligned measurement is still super-resolving and becomes optimal at $q=1/2$, when \(x_c=x_g\).

For completeness, we also consider the FI of the actual binary SPADE measurements centred at $x_c$ and $x_g$, whose 
probability distributions are given by \(\{p_0=\tr[\rho^{(1)}(\bm\theta)\ketbra{\psi(x_c)}{\psi(x_c)}], 
p_1=1-p_0\}\) and \(\{q_0=\tr[\rho^{(1)}(\bm\theta)\ketbra{\psi(x_g)}{\psi(x_g)}], q_1=1-q_0\}\) 
respectively. We find the same behaviour for the FI as that for the POVMs \(\{\ket{v_1^{(\hat{q})}}\bra{v_1^{(\hat{q})}},\, \ket{v_2^{(\hat{q})}}\bra{v_2^{(\hat{q})}}\}\) and \(\{\ketbra{e_1}{e_1},\ketbra{e_2}{e_2}\}\), respectively (in the limit $s/\sigma \ll 1$). The 
binary SPADE centred at $x_c$ gives the optimal FI, whereas the one located at $x_g$ gives the same sub-optimal FI. 
As previously, applying the qubit approximation to the latter measurement yields an 
overestimation of the QFI, which also coincides with naively including the FI of the binary SPADE measurement itself.

\subsection{\label{subs:qfi_indirect}Indirect estimation for the separation of the sources}

As we have shown, in the limit $s/\sigma\ll1$ we can safely restrict the imaging of partially coherent point 
sources to the qubit model where the pertinent parameters are encoded in the state's Bloch vector. An alternative 
method for extracting estimates of $\boldsymbol{\theta}$ proceeds via estimation of the Bloch vector $\mathbf{r}$ 
from which one can use the functional dependence on the parameters $\boldsymbol{\theta}$, and standard error 
propagation, to extract estimates for all pertinent parameters. For the case of incoherent sources, this 
indirect estimation method has been shown to furnish estimates of both the separation and 
relative intensity---by estimating the \emph{purity} $r_{\mathrm{inc}}$ of \eqnref{eq:r_incoh} and the relative 
intensity $q$---that saturate the QFI~\cite{deAlmeida21b}. We note that the resulting QFI for the separation agrees 
with that derived in \secref{subs:qfi_direct} with the intensity centroid, $x_0=x_c$, as the choice of reference frame 
(or equivalently, when $x_0$ is a nuisance parameter).
For the incoherent case~\cite{deAlmeida21b} show that such indirect estimation methods allow for simultaneous estimation 
of both the separation and relative intensity, saturating the multi-parameter Cram\'er-Rao bound, through collective measurement 
strategies.  In this section we ask whether such indirect estimation methods continue to be advantageous for the case of partially 
coherent sources. With the exception of $x_0$, which we assume known and equal to $x_c$, in what follows 
we are interested in estimating the parameters ${\bm\vartheta}=(r,q,\gamma_R,\gamma_I)$. 

To that end we proceed by estimating the purity (\eqnref{eq:r}) of the Bloch vector \(\mathbf{r}\), which itself 
carries information about $\bm\theta$. The SLD for the purity can be shown to be of the form 
$\Lambda_r = \lambda_r^{(0)}\openone + \boldsymbol{\lambda}_r \cdot \boldsymbol{\sigma}$, with
	\begin{equation}
    		\lambda_r^{(0)} = -\frac{r}{1-r^2} , \hspace{15pt} \boldsymbol{\lambda}_r = \frac{1}{1-r^2} 
            \frac{\mathbf{r}}{r} ,
	\label{eq:sld_r}
	\end{equation}
whilst the SLDs of $(q,\gamma_R,\gamma_I)$ keep their original expressions found in \cref{subs:qfi_direct}.  
The covariance matrix associated with \(\bm{\vartheta}=(r, q, \gamma_R,\gamma_I)^T\) can be calculated from the one 
associated to the original set of parameters $\boldsymbol{\theta}$ using
	\begin{equation}
    		\mathrm{Cov}(\bm\theta) =\mathcal{J} \mathrm{Cov}[\rho^{(1)}(\boldsymbol{\vartheta})] 
            \mathcal{J}^T \, ,
	\label{eq:QFI_through_purity}
	\end{equation}
where $\mathcal{J}$ is the Jacobian matrix of the transformation relating $\boldsymbol{\vartheta}$ with 
$\boldsymbol{\theta}$ whose elements are given by
	\begin{equation}
    		\mathcal{J}_{lk} = \frac{\partial \theta_k (\boldsymbol{\vartheta})}{\partial \vartheta_l}, 
            \hspace{10pt}\theta_k \in \boldsymbol{\theta}, \hspace{10pt} \vartheta_l \in \boldsymbol{\vartheta} .
	\label{eq:Jacobian_elements}
	\end{equation}

\begin{figure}
    		\centering
    		\includegraphics[width=0.8\columnwidth]{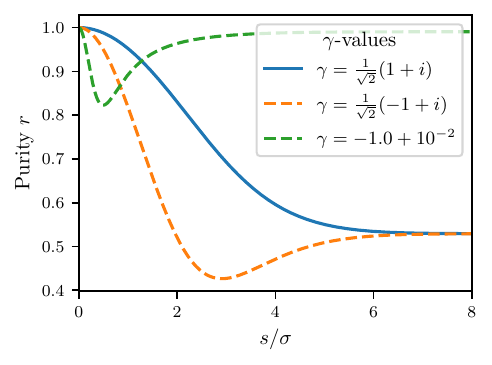}
    		\caption{Purity as a function of the separation for two values of $\gamma$: one with positive and one 
            with negative real part, as denoted by the legend. The plot is made for a relative intensity of 
            $q=0.4$. If $\gamma_R < 0$, the function $r(s)$ presents a local minimum, and the relation 
            $s \leftrightarrow r$ is not a bijection. In the limit $q\to1/2$, $\gamma_I\to0$ the local minimum 
            becomes sharp.}
    \label{fig:plot_r}
\end{figure}

It is important to note that this estimation strategy crucially depends on 
$\boldsymbol{\theta}\leftrightarrow\boldsymbol{\vartheta}$ being a bijection, which is not guaranteed. 
In fact, we find that
	\begin{equation}
    		\left.\frac{\partial r}{\partial c}\right|_{c=c_0} = 0 \hspace{1ex} \text{if} \hspace{1ex} 
            \gamma_R < 0,
 		\qquad c_0=-2\gamma_R \sqrt{q(1-q)},
	\label{eq:bijection_condition}
	\end{equation}
i.e., for negative $\gamma_R$ the purity presents a local minimum at $s_0 = \sqrt{-8\sigma^2 \ln c_0}$. 
An example of the serious implications posed by \eqnref{eq:bijection_condition} is shown in \cref{fig:plot_r}. 
Thus, such indirect strategies for estimating precision via estimation of the purity can only be applied if we know
\textit{a priori} that $\gamma_R \geq 0$. If $\gamma_R<0$, one could apply such indirect estimation so long as
 \(r > r_\infty\), where 
	\begin{equation}
    		r^2_\infty \coloneqq \lim_{s\to\infty}r^2 =1 + 4q(1-q)(\gamma_R^2+\gamma_I^2-1).
	\label{eq:condition}
	\end{equation}
However, knowing whether this bound is satisfied requires previous knowledge of $\gamma_I$, which is not accessible from the $s/\sigma \ll 1$ limit.
	
\begin{figure}
    \centering
    \includegraphics[width=0.8\columnwidth]{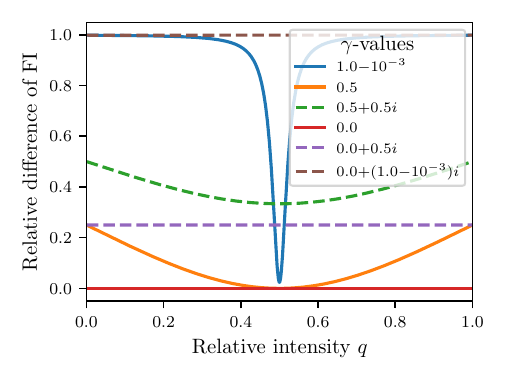}
    \caption{Relative difference for separation estimation between the direct estimation approach and the 
    indirect approach via purity estimation. The larger the difference the more inefficient the estimation through the purity becomes.}
    \label{fig:plot_qfir}
\end{figure}

Even if $\gamma_R \geq 0$, indirect estimation of the separation through the purity of the Bloch vector is 
sub-optimal for the case of partially coherent sources. In \cref{fig:plot_qfir} we compare the QFI matrix diagonal 
element for the separation, both for the direct and indirect strategies by means of their relative difference, showing 
that such an indirect strategy is optimal only in the case of incoherent sources, i.e., $\gamma=0$. For non-zero coherence 
factors, the efficiency of this strategy decreases with both $|\gamma_R|$ and $|\gamma_I|$. Consequently, the precision 
$H_s$ of the indirect strategy is also sub-optimal, even in the cases where other parameters are assumed to be known.

\section{\label{sec:conclusions} Conclusions}

This work studies the quantum multi-parameter estimation problem of imaging two partially coherent, 
quasi-monochromatic point sources.  We obtain the quantum mechanically optimal precisions for estimating 
the separation, relative intensity and coherence factor of the two point sources both in the case 
where some parameters are known or treated as nuisances.  Contrary to previous claims of a resurgence of 
Rayleigh's curse, we show that super-resolution is possible in both cases, with the  
degree of super-resolution depending strongly on both the real and imaginary part of the coherence factor.

We also study the efficacy of the qubit model---a two-dimensional simplification of the description of the PSF at
the image plane---applicable in the sub-diffraction regime.  We show that this model is applicable on the proviso 
that the PSFs of the two point sources, as well as the corresponding measurements,  are described relative to 
the same reference position $x_0$.  Finally, we also show that indirect estimation of the separation by first 
estimating the purity of the corresponding density matrix describing the PSF in the qubit model, is sub-optimal 
in the case of non-zero coherence factors and can even be ill-defined if $\gamma_R<0$.   

Our findings can be of practical relevance in a variety of imaging applications where partially coherent emitters appear 
such as fluorescence microscopy~\cite{Schwartz2013, Li2022}, as well as in the fabrication of photonic quantum devices 
such as quantum dots coupled to nanowires~\cite{Reitzenstein2006,Laussy2008,Laucht2009,Laussy2009,Laucht2010,Laussy2011,Lodahl2015,Utzat2019, Liu2024, Vasinka2025,Huang2025}. Note that in such scenarios the average photon number $\delta$ can be known ahead of time, since as 
the quantum dots are fabricated and characterized prior to being coupled with cavities or nanowires. The inclusion of the average photon number $\delta$ in the parameters to be estimated, the presence of background noise, resolution of multiple, partially coherent 
sources, as well as resolution of partially coherent sources in all three-dimensions---particularly axial resolution---are interesting directions for future investigation.

\section*{Acknowledgments} We would like to thank the anonymous reviewer for their invaluable suggestions. This work is funded through
Project PID2024-162141OB-I00, funded by MI-CIU/AEI/10.13039/501100011033/ FEDER, UE, Ayuda Ram\'on y Cajal 2021 (RYC2021-032032-I,
MICIU/AEI/10.13039/501100011033, ESF+) as well as Project FEDER C-EXP-256-UGR23 Consejer\'ia de Universidad, Investigaci\'on e Innovaci\'on y UE Programa FEDER Andaluc\'ia 2021-2027.


\renewcommand{\thesection}{Appendix}
\section{\label{app:SLD_computation}Computation of Symmetric Logarithmic Derivatives}

In this Supplement we provide the explicit computation for obtaining the SLD operators associated with each of the 
parameters $\boldsymbol{\theta}$.  Choosing our reference position $x_0$ as in \cref{eq:reference_position} in the main text, the 
corresponding quantum mechanical wavefunctions describing the state of a boson emanating from each of the 
sources are given by 
	\be
		\begin{split}
			\ket{\psi} &= \left(\frac{1}{2\pi\sigma^2}\right)^{1/4}\int_{-\infty}^{\infty} e^{-\frac{(x - x_0 +(1-
            \alpha)s)^2}{4\sigma^2}}\, \ket{x}\, \d x\\
			\ket{\phi} &= \left(\frac{1}{2\pi\sigma^2}\right)^{1/4}\int_{-\infty}^{\infty} e^{-\frac{(x - x_0 -\alpha s)^2}
            {4\sigma^2}}\, \ket{x}\, \d x\, .
		\end{split}
	\label{app:PSFs}
	\ee
Note that both wave functions depend explicitly on the reference frame position $x_0$ and the separation $s$, but not on any of the other parameters 
$q$, $\gamma_R$ and $\gamma_I$. For the latter three parameters the SLDs are given by (using \cref{eq:SLD} in the main text)  
	\begin{align}
        \lambda_q^{(0)} &= \frac{1-2q}{A}, \quad 
        \boldsymbol{\lambda}_q= \frac{1}{A} \begin{pmatrix}
            \sqrt{1-c^2} \\ 0 \\ c(1-2q)
        \end{pmatrix}\, , \\ \nonumber
    		\lambda_{\gamma_R}^{(0)} &= -\frac{2c(1-\gamma_I^2)\sqrt{q(1-q)} + \gamma_R}{B}\, , \\
		\lambda_{\gamma_I}^{(0)} &= -\frac{\gamma_I}{1 - \gamma_R^2 - \gamma_I^2}\, , \\ \nonumber
    		\boldsymbol{\lambda}_{\gamma_R} &= -\frac{1}{B}\begin{pmatrix}
        											\gamma_R(1-2q)\sqrt{1-c^2} \\
        											\frac{2\gamma_R\gamma_I\sqrt{1-c^2} \left( 2c\gamma_Rq(1-q) 
                                                    + \sqrt{q(1-q)} \right)}
        											{\left( 1 + 2c \gamma_R\sqrt{q(1-q)} \right)} \\
        											c\gamma_R + 2(1-\gamma_I^2)\sqrt{q(1-q)}
    											\end{pmatrix}\, ,\\
    		\boldsymbol{\lambda}_{\gamma_I} &=\frac{1}{B}\begin{pmatrix}
        											\gamma_I (2q-1) \sqrt{1-c^2} \\
        											-2(1-\gamma_R^2) \sqrt{1-c^2} \sqrt{q(1-q)} \\
        											-\gamma_I \left( c + 2\gamma_R \sqrt{q(1-q)} \right)
    											\end{pmatrix}\, ,
	\label{app:SLD_gammas}
	\end{align}
where $A = 2q(1-q)\left(1+2c\gamma_R\sqrt{q(1-q)}\right)$ and \(B=(1-\gamma_R^2-\gamma_I^2)\left( 1 + 2c 
\gamma_R\sqrt{q(1-q)} \right)\).

The determination of the SLD for $s$ requires a bit more work since as the states \(\ket{e_1}, \ket{e_2}\) of 
Eq.(27) in the main text depend explicitly on $s$.  This means that we must also consider the 
states $\ket{\partial_s e_1}, \ket{\partial_s e_2}$---which do not belong to the subspace $\cH^{(1)}$---and we 
must consider an extended space, $\cH$, spanned by \(\{\ket{e_1}, \ket{e_2}, \ket{\partial_s e_1}, 
\ket{\partial_s e_2}\}\).

In order to construct the appropriate SLD we must first construct an orthonormal basis for $\cH$.  To that end we
define 
	\be
		\begin{split}
			\beta &= -\frac{sc}{8\sigma^2}\\
			\zeta &= \frac{(s^2-4\sigma^2)c}{64\sigma^4}\, ,
		\end{split} 
	\label{app:beta_y_zeta}
	\ee
so that 
	\begin{align}\nonumber
		\inner{\psi}{\partial_s\phi} & = 2\alpha\beta\\ \nonumber
		\inner{\phi}{\partial_s\psi} & = 2(1-\alpha)\beta\\  \nonumber
		\inner{\partial_s\psi}{\partial_s\psi} &= \frac{(1-\alpha)^2}{4\sigma^2}\\ \nonumber
		\inner{\partial_s\phi}{\partial_s\phi} &= \frac{\alpha^2}{4\sigma^2}\\
		\inner{\partial_s\psi}{\partial_s\phi} &= 4\alpha(1-\alpha)\zeta\, .
	\label{app:overlaps_states_ders}
	\end{align}
Using the Gram-Schmidt procedure we construct two additional orthonormal vectors 
	\begin{align}\nonumber
		\ket{e_3}&= \frac{1}{\sqrt{\omega^2_1-\nu^2}}\left(\ket{\partial_s e_1}-\nu\ket{e_2}\right)\\ \nonumber
		\ket{e_4}&=\frac{1}{\sqrt{\omega^2_2-\nu^2-\frac{\mu^2}{\omega^2_1-\nu^2}}}\Big(\ket{\partial_s 
        e_2}+\nu\ket{e_1}\\
		&-\frac{\mu}{\sqrt{\omega^2_1-\nu^2}}\ket{e_3}\Big)
	\label{app:e3ye4}
	\end{align}
where
	\begin{align}\nonumber
		\omega^2_1&=\frac{1}{1-c}\left(\frac{\alpha^2+(1-\alpha)^2}{8\sigma^2}-4\alpha(1-\alpha)\zeta-
        \frac{\beta^2}{1-c}\right)\\  \nonumber
		\omega^2_2&=\frac{1}{1+c}\left(\frac{\alpha^2+(1-\alpha)^2}{8\sigma^2}+4\alpha(1-\alpha)\zeta-
        \frac{\beta^2}{1+c}\right)\\ \nonumber
		\mu &=\frac{(1-2\alpha)}{\sqrt{1-c^2}}\left(\frac{1}{8\sigma^2}-\frac{2\beta^2}{1-c^2}\right)\\
		\nu &= \frac{\beta(1-2\alpha)}{\sqrt{1-c^2}}=\inner{\partial_s e_1}{e_2}=-\inner{\partial_2 e_2}{e_1}\, .
	\label{app:constants}
    \end{align}

With respect to the orthonormal basis \(\{\ket{e_1},\, \ket{e_2}, \, \ket{e_3}, \, \ket{e_4} \}\) we can write 	
\begin{align}\nonumber
		\rho^{(1)}&=\left(\begin{array}{c|c}
					\rho^{(1)} & \mathbf{0}\\
					\hline
					\mathbf{0} & \mathbf{0}
				\end{array}\right)\\  \nonumber
		\frac{\partial\rho^{(1)}}{\partial s} &=\left(\begin{array}{c|c} 
        							     A^{(\text{qb})} + A^{(\text{ex})} & C \\ 
        								\hline 
        							    C^{\dagger} & \mathbf{0} 
    							    \end{array} \right) \\
    		\Lambda_s &=\left(\begin{array}{c|c} 
        						\Lambda_{11} & \Lambda_{12} \\ 
        						\hline 
  						\Lambda_{12}^{\dagger} & \Lambda_{22} 
    					\end{array}\right)\, ,
	\label{app:SLD_s1}
	\end{align}
where \(\mathbf{0}\) denotes the two-dimensional zero matrix, \(A^{(\text{qb})},\, A^{(\text{ex})}, 
\Lambda_{11}:\cH^{(1)}\to\cH^{(1)}\), \(C,\Lambda_{12}:\cH^{(2)}\to\cH^{(1)} \) and 
\(\Lambda_{22}:\cH^{(2)}\to\cH^{(2)}\).  Here, \(\cH^{(1)}, \, \cH^{(2)}\) denote the subspaces spanned by 
\(\{\ket{e_1},\, \ket{e_2}\}\) and \(\{\ket{e_3},\, \ket{e_4}\}\) respectively.  Using \cref{eq:deriv_rho_s} in the main text 
the matrices \(A^{(\text{qb})},\, A^{(\text{ex})}, C\) are explicitly given as
	\begin{align}\nonumber
		&A^{(\text{qb})}=\frac{1}{2}\frac{\partial\mathbf{r}}{\partial s}\cdot \paulivec\\ \nonumber
		&A^{(\text{ex})}=\nu\begin{pmatrix}
   					-2 \Re (\rho_{12})& (\rho_{11} - \rho_{22})\\
        					(\rho_{11} - \rho_{22}) & 2 \Re (\rho_{12})
    					\end{pmatrix}\\
		&C = \small{\begin{pmatrix}
        				\rho_{11} \sqrt{\omega^2_1 - \nu^2} + \rho_{12} \frac{\mu}{\sqrt{\omega^2_1 - \nu^2}} &
        				\hspace{10pt} & \rho_{12} \sqrt{\omega^2_2 - \nu^2 - \frac{\mu^2}{\omega^2_1 - \nu^2}} \\
        				\\\rho_{21} \sqrt{\omega^2_1 - \nu^2} + \rho_{22} \frac{\mu}{\sqrt{\omega^2_1 - \nu^2}} &
        				\hspace{10pt} &\rho_{22} \sqrt{\omega^2_2 - \nu^2 - \frac{\mu^2}{\omega^2_1 - \nu^2}}
   			 \end{pmatrix}}
	\label{app:AsandC}
	\end{align}
where \(\Re(z)\) denotes the real part of the complex number \(z\).  

Substituting \eqnref{app:SLD_s1} into \cref{eq:SLD} of the main text yields the following set of equations for each of the 
two-by-two blocks 
	\begin{align}\nonumber
    		A^{(\text{qb})} + A^{(\text{ex})} &= \frac{\Lambda_{11}\rho^{(1)} + \rho^{(1)}\Lambda_{11}}{2}\\ 
            \nonumber
    		C &= \frac{\rho^{(1)} \Lambda_{12}}{2}\\
    		\mathbf{0}_{2\times2} &= \Lambda_{22}\, .
	\label{app:block_solutions}
	\end{align}
The second identity in \eqnref{app:block_solutions} yields immediately the solution  
	\begin{equation} 
		\begin{split}
    			\Lambda_{12} &= 2 {\rho^{(1)}}^{-1} C \\
    			&= 2 \begin{pmatrix}\sqrt{\omega^2_1 - \nu^2} & 0 \\
        				\frac{\mu}{\sqrt{\omega^2_1 - \nu^2}} & \sqrt{\omega^2_2 - \nu^2 - \frac{\mu^2}
                        {\omega^2_1 - \nu^2}}\, .
    			\end{pmatrix}
		\end{split}
	\label{app:Lambda_12}
	\end{equation}
To find the solution for $\Lambda_{11}$, we can write the blocks involved in the first equation of 
\eqnref{app:AsandC} in terms of their Bloch form as
	\begin{align}\nonumber
    		A^{(\text{qb})} &= \frac{1}{2} \frac{\partial \mathbf{r}}{\partial s} \cdot \boldsymbol{\sigma}\\  
            \nonumber
    		A^{(\text{ex})} &= d^{(0)}\openone + \mathbf{d} \cdot \boldsymbol{\sigma}\\
    		\Lambda_{11} &= \lambda_s^{(0)} \openone + \boldsymbol{\lambda}_s \cdot \boldsymbol{\sigma}\, ,
	\label{app:bloch_form}
	\end{align}
with $\mathbf{r}$ given in \cref{eq:Bloch_vector} of the main text and
	\begin{align} \nonumber
    		d^{(0)} &= 0\\  
    		\mathbf{d} &=\frac{-\nu}{1 + 2c\gamma_R \sqrt{q(1-q)}}
    		\begin{pmatrix}c + 2\gamma_R \sqrt{q(1-q)} \\0 \\(2q-1) \sqrt{1-c^2}
    		\end{pmatrix}\, .
	\label{app:d}
	\end{align}
Expressing the blocks this way allows for an easy procedure to find
	\begin{equation}
    		\begin{split}
		\lambda_s^{(0)} &= \frac{\left( \frac{\partial \mathbf{r}}{\partial s} \cdot \mathbf{r} \right)
    						+ 2 \left( \mathbf{d} \cdot \mathbf{r} - d^{(0)} \right) }{r^2 - 1}\\
    		\boldsymbol{\lambda}_s &= \frac{\partial \mathbf{r}}{\partial s} + 2 \mathbf{d} - \lambda_s^{(0)} 
            \mathbf{r}\, .
		\end{split}
	\label{app:lambda_s}
	\end{equation}
Thus, we can decompose $\Lambda_{11}$ into a sum of two contributions: a ``qubit'' term and an ``extra'' one:
	\begin{equation}
    		\begin{split}
			\Lambda_{11} &= \Lambda_{11}^{(\text{qb})} + \Lambda_{11}^{(\text{ex})}\\
			\Lambda_{11}^{(\text{qb})} &= \lambda^{(0,\text{qb})}_{s} \openone + 
            \boldsymbol{\lambda}_{s}^{(\text{qb})}\cdot\boldsymbol{\sigma}\\
   			\Lambda_{11}^{(\text{ex})} &= \lambda^{(0,\text{ex})}_{s} \openone + 
            \boldsymbol{\lambda}_{s}^{(\text{ex})}\cdot\boldsymbol{\sigma}\, ,
		\end{split}
	\label{app:Lambda_11}
	\end{equation}
whose Bloch decompositions are explicit given by
	\begin{align}\nonumber
    		\lambda^{(0,\text{qb})}_{s} &= \frac{\frac{\partial\mathbf{r}}{\partial s} \cdot \mathbf{r}}{r^2 - 1} 
            =\frac{-2\beta}{(1-c^2)}\left(\frac{c+2\gamma_R\sqrt{q(1-q)}}
    		{\left(1+2c\gamma_R \sqrt{q(1-q)} \right)}\right)\\ \nonumber
     		\lambda^{(0,\text{ex})}_{s} &= 2\frac{\mathbf{d}\cdot\mathbf{r} - d^{(0)}}{r^2-1} = 0\\ \nonumber
    		\boldsymbol{\lambda}_{s}^{(\text{qb})} &= \frac{\partial\mathbf{r}}{\partial s} - 
            \lambda^{(0,\text{qb})}_{s} \mathbf{r}
    		=\frac{-2\beta}{1-c^2}\begin{pmatrix}0 \\ 0 \\ 1\end{pmatrix}\\ \nonumber
    		\boldsymbol{\lambda}_{s}^{(\text{ex})} &= 2 \mathbf{d} - \lambda^{(0,\text{ex})}_{s} \mathbf{r} \\\
    		&=\frac{-2\nu}{1 + 2c\gamma_R \sqrt{q(1-q)}}\begin{pmatrix}c + 2\gamma_R \sqrt{q(1-q)} \\0 \\(2q-1) 
            \sqrt{1-c^2}\end{pmatrix}\, .
	\label{app:Lambda_11explicit}
	\end{align}

The SLD corresponding to the reference frame position, $\Lambda_{x_0}$, is also 4-dimensional, acting over the space spanned by $\{\ket{e_1},\ket{e_2},\ket{e_3},\ket{e_4} \}$. It is given by \cref{eq:SLD} of the main manuscript, with the derivative of $\rho^{(1)}(\bm\theta)$ being
\begin{equation}
    \frac{\partial\rho^{(1)}}{\partial x_0} =
    \sum_{i,j=1}^2 \rho_{ij} \Big( \ket{\partial_{x_0} e_i}\bra{e_j} +
    \ket{e_i}\bra{\partial_{x_0} e_j} \Big) .
\end{equation}

The derivatives of the quantum states $\ket{e_i}$ are in turn given by
\begin{align}
    \ket{\partial_{x_0} e_1} &=
    \frac{\beta(2\alpha-1) + \nu(1-c)\sqrt{\frac{1+c}{1-c}}}{2\alpha(1-\alpha)(1-c)}\ket{e_1} \nonumber \\
    &\qquad -
    \frac{\frac{\beta}{\sqrt{1-c^2}}+\nu(2\alpha-1)}{2\alpha(1-\alpha)} \ket{e_2} \nonumber\\
    &\qquad
    - \frac{\mu\sqrt{\frac{1+c}{1-c}}+(2\alpha-1)(\omega_1^2-\nu^2)}{2\alpha(1-\alpha)\sqrt{\omega_1^2-\nu^2}} \ket{e_3} \nonumber \\
    &\qquad -
    \frac{\sqrt{\frac{1+c}{1-c}}\sqrt{\omega_2^2-\nu^2-\frac{\mu^2}{\omega_1^2-\nu^2}}}{2\alpha(1-\alpha)}\ket{e_4} \,,
    \end{align}
    \begin{align}
    \ket{\partial_{x_0}e_2} &= 
    \frac{\frac{\beta}{\sqrt{1-c^2}}+\nu(2\alpha-1)}{2\alpha(1-\alpha)} \ket{e_1} \nonumber\\
    &\quad -
    \frac{\frac{1}{\sqrt{1-c}}\left(
        \beta (2\alpha-1)\sqrt{\frac{2}{1+c}-1} +\nu(1-c)
    \right)}{2\alpha(1-\alpha)\sqrt{1-c}} \ket{e_2} \nonumber \\
    &\quad -
    \frac{\mu(2\alpha-1)\sqrt{1-c} + \frac{1}{\sqrt{1+c}}(1-c)(\omega_1^2-\nu^2)}{2\alpha(1-\alpha)\sqrt{1-c}\sqrt{\omega_1^2-\nu^2}} \ket{e_3} \nonumber \\
    &\quad
    - \frac{(2\alpha-1)\sqrt{\omega_2^2 - \nu^2 - \frac{\mu^2}{\omega_1^2-\nu^2}}}{2\alpha(1-\alpha)} \ket{e_4} \,.
\end{align}
Form here the procedure to compute $\Lambda_{x_0}$ is completely analogous to that of $\Lambda_s$. Although numerically well-behaved, the expression of this SLD is too cumbersome and not particularly illuminating, hence we have opted not to include it in this paper.

Given the explicit form for the SLDs we can now easily check their commutation relations. To the smallest 
non-trivial order in $s$ we find $[\Lambda_s,\Lambda_q]$, $[\Lambda_s,\Lambda_{\gamma_R}]$,
$[\Lambda_s,\Lambda_{\gamma_I}]$, $[\Lambda_{x_0},\Lambda_q]$, $[\Lambda_{x_0},\Lambda_{\gamma_R}]$, $[\Lambda_{x_0},\Lambda_{\gamma_I}]$ are all $\cO(1)$---independent of $s$---whilst $[\Lambda_q,\Lambda_{\gamma_R}],
[\Lambda_q,\Lambda_{\gamma_I}],[\Lambda_{\gamma_R},\Lambda_{\gamma_I}]$ all have leading order $\cO(s)$. Lastly, $[\Lambda_{x_0},\Lambda_s]$ has leading order $\cO(s^{-1})$.
For the necessary and sufficient condition of \cref{eq:QCRB_attainability} of the main paper, we find instead $\tr[\rho^{(1)}\left[\Lambda_{x_0}, \Lambda_s]\right]$ has leading order $\cO(1)$, $\tr\left[\rho^{(1)}
[\Lambda_s,\Lambda_q]\right]$, $\tr\left[\rho^{(1)}[\Lambda_s,\Lambda_{\gamma_R}]\right]$, 
$\tr\left[\rho^{(1)}[\Lambda_s,\Lambda_{\gamma_I}]\right]$, $\tr[\rho^{(1)}\left[\Lambda_{x_0}, \Lambda_q]\right]$, $\tr[\rho^{(1)}\left[\Lambda_{x_0}, \Lambda_{\gamma_R}]\right]$, $\tr[\rho^{(1)}\left[\Lambda_{x_0}, \Lambda_{\gamma_I}]\right]$ all have leading order $\cO(s)$, whilst 
$\tr\left[\rho^{(1)}[\Lambda_q,\Lambda_{\gamma_R}]\right]$, 
$\tr\left[\rho^{(1)}[\Lambda_q,\Lambda_{\gamma_I}]\right]$, 
$\tr\left[\rho^{(1)}[\Lambda_{\gamma_I},\Lambda_{\gamma_R}]\right]$ all 
have leading order $\cO(s^2)$.

\bibliographystyle{apsrev4-2}
\bibliography{superres}

\begin{thebibliography}{58}%
\makeatletter
\providecommand \@ifxundefined [1]{%
 \@ifx{#1\undefined}
}%
\providecommand \@ifnum [1]{%
 \ifnum #1\expandafter \@firstoftwo
 \else \expandafter \@secondoftwo
 \fi
}%
\providecommand \@ifx [1]{%
 \ifx #1\expandafter \@firstoftwo
 \else \expandafter \@secondoftwo
 \fi
}%
\providecommand \natexlab [1]{#1}%
\providecommand \enquote  [1]{``#1''}%
\providecommand \bibnamefont  [1]{#1}%
\providecommand \bibfnamefont [1]{#1}%
\providecommand \citenamefont [1]{#1}%
\providecommand \href@noop [0]{\@secondoftwo}%
\providecommand \href [0]{\begingroup \@sanitize@url \@href}%
\providecommand \@href[1]{\@@startlink{#1}\@@href}%
\providecommand \@@href[1]{\endgroup#1\@@endlink}%
\providecommand \@sanitize@url [0]{\catcode `\\12\catcode `\$12\catcode `\&12\catcode `\#12\catcode `\^12\catcode `\_12\catcode `\%12\relax}%
\providecommand \@@startlink[1]{}%
\providecommand \@@endlink[0]{}%
\providecommand \url  [0]{\begingroup\@sanitize@url \@url }%
\providecommand \@url [1]{\endgroup\@href {#1}{\urlprefix }}%
\providecommand \urlprefix  [0]{URL }%
\providecommand \Eprint [0]{\href }%
\providecommand \doibase [0]{https://doi.org/}%
\providecommand \selectlanguage [0]{\@gobble}%
\providecommand \bibinfo  [0]{\@secondoftwo}%
\providecommand \bibfield  [0]{\@secondoftwo}%
\providecommand \translation [1]{[#1]}%
\providecommand \BibitemOpen [0]{}%
\providecommand \bibitemStop [0]{}%
\providecommand \bibitemNoStop [0]{.\EOS\space}%
\providecommand \EOS [0]{\spacefactor3000\relax}%
\providecommand \BibitemShut  [1]{\csname bibitem#1\endcsname}%
\let\auto@bib@innerbib\@empty
\bibitem [{\citenamefont {Rayleigh}(1880)}]{Rayleigh80}%
  \BibitemOpen
  \bibfield  {author} {\bibinfo {author} {\bibfnamefont {L.}~\bibnamefont {Rayleigh}},\ }\href@noop {} {\bibfield  {journal} {\bibinfo  {journal} {Lond.Edinb.Dubl.Phil.Mag.}\ }\textbf {\bibinfo {volume} {10}},\ \bibinfo {pages} {116} (\bibinfo {year} {1880})}\BibitemShut {NoStop}%
\bibitem [{\citenamefont {Tsang}\ \emph {et~al.}(2016{\natexlab{a}})\citenamefont {Tsang}, \citenamefont {Nair},\ and\ \citenamefont {Lu}}]{Tsang16}%
  \BibitemOpen
  \bibfield  {author} {\bibinfo {author} {\bibfnamefont {M.}~\bibnamefont {Tsang}}, \bibinfo {author} {\bibfnamefont {R.}~\bibnamefont {Nair}},\ and\ \bibinfo {author} {\bibfnamefont {X.-M.}\ \bibnamefont {Lu}},\ }\href {https://doi.org/10.1103/PhysRevX.6.031033} {\bibfield  {journal} {\bibinfo  {journal} {Phys. Rev. X}\ }\textbf {\bibinfo {volume} {6}},\ \bibinfo {pages} {031033} (\bibinfo {year} {2016}{\natexlab{a}})}\BibitemShut {NoStop}%
\bibitem [{\citenamefont {Tsang}\ \emph {et~al.}(2016{\natexlab{b}})\citenamefont {Tsang}, \citenamefont {Nair},\ and\ \citenamefont {Lu}}]{Tsang16b}%
  \BibitemOpen
  \bibfield  {author} {\bibinfo {author} {\bibfnamefont {M.}~\bibnamefont {Tsang}}, \bibinfo {author} {\bibfnamefont {R.}~\bibnamefont {Nair}},\ and\ \bibinfo {author} {\bibfnamefont {X.-M.}\ \bibnamefont {Lu}},\ }in\ \href {https://doi.org/10.1117/12.2245733} {\emph {\bibinfo {booktitle} {Quantum and Nonlinear Optics IV}}},\ Vol.\ \bibinfo {volume} {10029},\ \bibinfo {editor} {edited by\ \bibinfo {editor} {\bibfnamefont {Q.}~\bibnamefont {Gong}}, \bibinfo {editor} {\bibfnamefont {G.-C.}\ \bibnamefont {Guo}},\ and\ \bibinfo {editor} {\bibfnamefont {B.~S.}\ \bibnamefont {Ham}}},\ \bibinfo {organization} {International Society for Optics and Photonics}\ (\bibinfo  {publisher} {SPIE},\ \bibinfo {year} {2016})\ p.\ \bibinfo {pages} {1002903}\BibitemShut {NoStop}%
\bibitem [{\citenamefont {Nair}\ and\ \citenamefont {Tsang}(2016)}]{Nair16}%
  \BibitemOpen
  \bibfield  {author} {\bibinfo {author} {\bibfnamefont {R.}~\bibnamefont {Nair}}\ and\ \bibinfo {author} {\bibfnamefont {M.}~\bibnamefont {Tsang}},\ }\href {https://doi.org/10.1364/OE.24.003684} {\bibfield  {journal} {\bibinfo  {journal} {Opt. Express}\ }\textbf {\bibinfo {volume} {24}},\ \bibinfo {pages} {3684} (\bibinfo {year} {2016})}\BibitemShut {NoStop}%
\bibitem [{\citenamefont {Lupo}\ and\ \citenamefont {Pirandola}(2016)}]{Lupo16}%
  \BibitemOpen
  \bibfield  {author} {\bibinfo {author} {\bibfnamefont {C.}~\bibnamefont {Lupo}}\ and\ \bibinfo {author} {\bibfnamefont {S.}~\bibnamefont {Pirandola}},\ }\href {https://doi.org/10.1103/PhysRevLett.117.190802} {\bibfield  {journal} {\bibinfo  {journal} {Phys. Rev. Lett.}\ }\textbf {\bibinfo {volume} {117}},\ \bibinfo {pages} {190802} (\bibinfo {year} {2016})}\BibitemShut {NoStop}%
\bibitem [{\citenamefont {Ang}\ \emph {et~al.}(2017)\citenamefont {Ang}, \citenamefont {Nair},\ and\ \citenamefont {Tsang}}]{Ang17}%
  \BibitemOpen
  \bibfield  {author} {\bibinfo {author} {\bibfnamefont {S.~Z.}\ \bibnamefont {Ang}}, \bibinfo {author} {\bibfnamefont {R.}~\bibnamefont {Nair}},\ and\ \bibinfo {author} {\bibfnamefont {M.}~\bibnamefont {Tsang}},\ }\href {https://doi.org/10.1103/PhysRevA.95.063847} {\bibfield  {journal} {\bibinfo  {journal} {Phys. Rev. A}\ }\textbf {\bibinfo {volume} {95}},\ \bibinfo {pages} {063847} (\bibinfo {year} {2017})}\BibitemShut {NoStop}%
\bibitem [{\citenamefont {{\v{R}}eha{\v{c}}ek}\ \emph {et~al.}(2017)\citenamefont {{\v{R}}eha{\v{c}}ek}, \citenamefont {Hradil}, \citenamefont {Stoklasa}, \citenamefont {Pa{\'u}r}, \citenamefont {Grover}, \citenamefont {Krzic},\ and\ \citenamefont {S{\'a}nchez-Soto}}]{Rehacek17}%
  \BibitemOpen
  \bibfield  {author} {\bibinfo {author} {\bibfnamefont {J.}~\bibnamefont {{\v{R}}eha{\v{c}}ek}}, \bibinfo {author} {\bibfnamefont {Z.}~\bibnamefont {Hradil}}, \bibinfo {author} {\bibfnamefont {B.}~\bibnamefont {Stoklasa}}, \bibinfo {author} {\bibfnamefont {M.}~\bibnamefont {Pa{\'u}r}}, \bibinfo {author} {\bibfnamefont {J.}~\bibnamefont {Grover}}, \bibinfo {author} {\bibfnamefont {A.}~\bibnamefont {Krzic}},\ and\ \bibinfo {author} {\bibfnamefont {L.~L.}\ \bibnamefont {S{\'a}nchez-Soto}},\ }\href {https://doi.org/10.1103/PhysRevA.96.062107} {\bibfield  {journal} {\bibinfo  {journal} {Phys. Rev. A}\ }\textbf {\bibinfo {volume} {96}},\ \bibinfo {pages} {062107} (\bibinfo {year} {2017})}\BibitemShut {NoStop}%
\bibitem [{\citenamefont {{\v{R}}eh{\'a}{\v{c}}ek}\ \emph {et~al.}(2018)\citenamefont {{\v{R}}eh{\'a}{\v{c}}ek}, \citenamefont {Hradil}, \citenamefont {Koutn{\`y}}, \citenamefont {Grover}, \citenamefont {Krzic},\ and\ \citenamefont {S{\'a}nchez-Soto}}]{Rehacek18}%
  \BibitemOpen
  \bibfield  {author} {\bibinfo {author} {\bibfnamefont {J.}~\bibnamefont {{\v{R}}eh{\'a}{\v{c}}ek}}, \bibinfo {author} {\bibfnamefont {Z.}~\bibnamefont {Hradil}}, \bibinfo {author} {\bibfnamefont {D.}~\bibnamefont {Koutn{\`y}}}, \bibinfo {author} {\bibfnamefont {J.}~\bibnamefont {Grover}}, \bibinfo {author} {\bibfnamefont {A.}~\bibnamefont {Krzic}},\ and\ \bibinfo {author} {\bibfnamefont {L.~L.}\ \bibnamefont {S{\'a}nchez-Soto}},\ }\href {https://doi.org/10.1103/PhysRevA.98.012103} {\bibfield  {journal} {\bibinfo  {journal} {Phys. Rev. A}\ }\textbf {\bibinfo {volume} {98}},\ \bibinfo {pages} {012103} (\bibinfo {year} {2018})}\BibitemShut {NoStop}%
\bibitem [{\citenamefont {Lu}\ \emph {et~al.}(2018)\citenamefont {Lu}, \citenamefont {Krovi}, \citenamefont {Nair}, \citenamefont {Guha},\ and\ \citenamefont {Shapiro}}]{Lu18}%
  \BibitemOpen
  \bibfield  {author} {\bibinfo {author} {\bibfnamefont {X.-M.}\ \bibnamefont {Lu}}, \bibinfo {author} {\bibfnamefont {H.}~\bibnamefont {Krovi}}, \bibinfo {author} {\bibfnamefont {R.}~\bibnamefont {Nair}}, \bibinfo {author} {\bibfnamefont {S.}~\bibnamefont {Guha}},\ and\ \bibinfo {author} {\bibfnamefont {J.~H.}\ \bibnamefont {Shapiro}},\ }\href {https://doi.org/https://doi.org/10.1038/s41534-018-0114-y} {\bibfield  {journal} {\bibinfo  {journal} {npj Quantum Information}\ }\textbf {\bibinfo {volume} {4}},\ \bibinfo {pages} {64} (\bibinfo {year} {2018})}\BibitemShut {NoStop}%
\bibitem [{\citenamefont {Napoli}\ \emph {et~al.}(2019)\citenamefont {Napoli}, \citenamefont {Piano}, \citenamefont {Leach}, \citenamefont {Adesso},\ and\ \citenamefont {Tufarelli}}]{Napoli19}%
  \BibitemOpen
  \bibfield  {author} {\bibinfo {author} {\bibfnamefont {C.}~\bibnamefont {Napoli}}, \bibinfo {author} {\bibfnamefont {S.}~\bibnamefont {Piano}}, \bibinfo {author} {\bibfnamefont {R.}~\bibnamefont {Leach}}, \bibinfo {author} {\bibfnamefont {G.}~\bibnamefont {Adesso}},\ and\ \bibinfo {author} {\bibfnamefont {T.}~\bibnamefont {Tufarelli}},\ }\href {https://doi.org/10.1103/PhysRevLett.122.140505} {\bibfield  {journal} {\bibinfo  {journal} {Phys. Rev. Lett.}\ }\textbf {\bibinfo {volume} {122}},\ \bibinfo {pages} {140505} (\bibinfo {year} {2019})}\BibitemShut {NoStop}%
\bibitem [{\citenamefont {Prasad}\ and\ \citenamefont {Yu}(2019)}]{Prasad19}%
  \BibitemOpen
  \bibfield  {author} {\bibinfo {author} {\bibfnamefont {S.}~\bibnamefont {Prasad}}\ and\ \bibinfo {author} {\bibfnamefont {Z.}~\bibnamefont {Yu}},\ }\href {https://doi.org/10.1103/PhysRevA.99.022116} {\bibfield  {journal} {\bibinfo  {journal} {Phys. Rev. A}\ }\textbf {\bibinfo {volume} {99}},\ \bibinfo {pages} {022116} (\bibinfo {year} {2019})}\BibitemShut {NoStop}%
\bibitem [{\citenamefont {Tsang}(2019)}]{Tsang19}%
  \BibitemOpen
  \bibfield  {author} {\bibinfo {author} {\bibfnamefont {M.}~\bibnamefont {Tsang}},\ }\href {https://doi.org/10.1103/PhysRevA.99.012305} {\bibfield  {journal} {\bibinfo  {journal} {Phys. Rev. A}\ }\textbf {\bibinfo {volume} {99}},\ \bibinfo {pages} {012305} (\bibinfo {year} {2019})}\BibitemShut {NoStop}%
\bibitem [{\citenamefont {Lupo}\ \emph {et~al.}(2020)\citenamefont {Lupo}, \citenamefont {Huang},\ and\ \citenamefont {Kok}}]{Lupo20}%
  \BibitemOpen
  \bibfield  {author} {\bibinfo {author} {\bibfnamefont {C.}~\bibnamefont {Lupo}}, \bibinfo {author} {\bibfnamefont {Z.}~\bibnamefont {Huang}},\ and\ \bibinfo {author} {\bibfnamefont {P.}~\bibnamefont {Kok}},\ }\href {https://doi.org/10.1103/PhysRevLett.124.080503} {\bibfield  {journal} {\bibinfo  {journal} {Phys. Rev. Lett.}\ }\textbf {\bibinfo {volume} {124}},\ \bibinfo {pages} {080503} (\bibinfo {year} {2020})}\BibitemShut {NoStop}%
\bibitem [{\citenamefont {De~Almeida}\ \emph {et~al.}(2021)\citenamefont {De~Almeida}, \citenamefont {Ko{\l}ody{\'n}ski}, \citenamefont {Hirche}, \citenamefont {Lewenstein},\ and\ \citenamefont {Skotiniotis}}]{deAlmeida21}%
  \BibitemOpen
  \bibfield  {author} {\bibinfo {author} {\bibfnamefont {J.~O.}\ \bibnamefont {De~Almeida}}, \bibinfo {author} {\bibfnamefont {J.}~\bibnamefont {Ko{\l}ody{\'n}ski}}, \bibinfo {author} {\bibfnamefont {C.}~\bibnamefont {Hirche}}, \bibinfo {author} {\bibfnamefont {M.}~\bibnamefont {Lewenstein}},\ and\ \bibinfo {author} {\bibfnamefont {M.}~\bibnamefont {Skotiniotis}},\ }\href {https://doi.org/10.1103/PhysRevA.103.022406} {\bibfield  {journal} {\bibinfo  {journal} {Phys. Rev. A}\ }\textbf {\bibinfo {volume} {103}},\ \bibinfo {pages} {022406} (\bibinfo {year} {2021})}\BibitemShut {NoStop}%
\bibitem [{\citenamefont {Hu}\ \emph {et~al.}(2025)\citenamefont {Hu}, \citenamefont {Wang}, \citenamefont {Zhang}, \citenamefont {Wang}, \citenamefont {Liu}, \citenamefont {Zhou},\ and\ \citenamefont {Zhang}}]{Hu25}%
  \BibitemOpen
  \bibfield  {author} {\bibinfo {author} {\bibfnamefont {C.}~\bibnamefont {Hu}}, \bibinfo {author} {\bibfnamefont {B.}~\bibnamefont {Wang}}, \bibinfo {author} {\bibfnamefont {J.}~\bibnamefont {Zhang}}, \bibinfo {author} {\bibfnamefont {K.}~\bibnamefont {Wang}}, \bibinfo {author} {\bibfnamefont {H.}~\bibnamefont {Liu}}, \bibinfo {author} {\bibfnamefont {J.}~\bibnamefont {Zhou}},\ and\ \bibinfo {author} {\bibfnamefont {L.}~\bibnamefont {Zhang}},\ }\href {https://doi.org/10.1103/z157-q6n3} {\bibfield  {journal} {\bibinfo  {journal} {Phys. Rev. A}\ }\textbf {\bibinfo {volume} {112}},\ \bibinfo {pages} {032609} (\bibinfo {year} {2025})}\BibitemShut {NoStop}%
\bibitem [{\citenamefont {Katamadze}\ \emph {et~al.}(2025)\citenamefont {Katamadze}, \citenamefont {Bantysh}, \citenamefont {Chernyavskiy}, \citenamefont {Bogdanov},\ and\ \citenamefont {Kulik}}]{Katamadze25}%
  \BibitemOpen
  \bibfield  {author} {\bibinfo {author} {\bibfnamefont {K.}~\bibnamefont {Katamadze}}, \bibinfo {author} {\bibfnamefont {B.}~\bibnamefont {Bantysh}}, \bibinfo {author} {\bibfnamefont {A.}~\bibnamefont {Chernyavskiy}}, \bibinfo {author} {\bibfnamefont {Y.}~\bibnamefont {Bogdanov}},\ and\ \bibinfo {author} {\bibfnamefont {S.}~\bibnamefont {Kulik}},\ }\href {https://doi.org/10.1103/PhysRevApplied.23.024066} {\bibfield  {journal} {\bibinfo  {journal} {Phys. Rev. Appl.}\ }\textbf {\bibinfo {volume} {23}},\ \bibinfo {pages} {024066} (\bibinfo {year} {2025})}\BibitemShut {NoStop}%
\bibitem [{\citenamefont {Yang}\ \emph {et~al.}(2016)\citenamefont {Yang}, \citenamefont {Tashchilina}, \citenamefont {Moiseev}, \citenamefont {Simon},\ and\ \citenamefont {Lvovsky}}]{Yang16}%
  \BibitemOpen
  \bibfield  {author} {\bibinfo {author} {\bibfnamefont {F.}~\bibnamefont {Yang}}, \bibinfo {author} {\bibfnamefont {A.}~\bibnamefont {Tashchilina}}, \bibinfo {author} {\bibfnamefont {E.~S.}\ \bibnamefont {Moiseev}}, \bibinfo {author} {\bibfnamefont {C.}~\bibnamefont {Simon}},\ and\ \bibinfo {author} {\bibfnamefont {A.~I.}\ \bibnamefont {Lvovsky}},\ }\href {https://doi.org/10.1364/OPTICA.3.001148} {\bibfield  {journal} {\bibinfo  {journal} {Optica}\ }\textbf {\bibinfo {volume} {3}},\ \bibinfo {pages} {1148} (\bibinfo {year} {2016})}\BibitemShut {NoStop}%
\bibitem [{\citenamefont {Tham}\ \emph {et~al.}(2017)\citenamefont {Tham}, \citenamefont {Ferretti},\ and\ \citenamefont {Steinberg}}]{Tham17}%
  \BibitemOpen
  \bibfield  {author} {\bibinfo {author} {\bibfnamefont {W.-K.}\ \bibnamefont {Tham}}, \bibinfo {author} {\bibfnamefont {H.}~\bibnamefont {Ferretti}},\ and\ \bibinfo {author} {\bibfnamefont {A.~M.}\ \bibnamefont {Steinberg}},\ }\href {https://doi.org/10.1103/PhysRevLett.118.070801} {\bibfield  {journal} {\bibinfo  {journal} {Phys. Rev. Lett.}\ }\textbf {\bibinfo {volume} {118}},\ \bibinfo {pages} {070801} (\bibinfo {year} {2017})}\BibitemShut {NoStop}%
\bibitem [{\citenamefont {Yang}\ \emph {et~al.}(2017)\citenamefont {Yang}, \citenamefont {Nair}, \citenamefont {Tsang}, \citenamefont {Simon},\ and\ \citenamefont {Lvovsky}}]{Yang17}%
  \BibitemOpen
  \bibfield  {author} {\bibinfo {author} {\bibfnamefont {F.}~\bibnamefont {Yang}}, \bibinfo {author} {\bibfnamefont {R.}~\bibnamefont {Nair}}, \bibinfo {author} {\bibfnamefont {M.}~\bibnamefont {Tsang}}, \bibinfo {author} {\bibfnamefont {C.}~\bibnamefont {Simon}},\ and\ \bibinfo {author} {\bibfnamefont {A.~I.}\ \bibnamefont {Lvovsky}},\ }\href {https://doi.org/10.1103/PhysRevA.96.063829} {\bibfield  {journal} {\bibinfo  {journal} {Phys. Rev. A}\ }\textbf {\bibinfo {volume} {96}},\ \bibinfo {pages} {063829} (\bibinfo {year} {2017})}\BibitemShut {NoStop}%
\bibitem [{\citenamefont {Donohue}\ \emph {et~al.}(2018)\citenamefont {Donohue}, \citenamefont {Ansari}, \citenamefont {\ifmmode \check{R}\else \v{R}\fi{}eh\'a\ifmmode~\check{c}\else \v{c}\fi{}ek}, \citenamefont {Hradil}, \citenamefont {Stoklasa}, \citenamefont {Pa\'ur}, \citenamefont {S\'anchez-Soto},\ and\ \citenamefont {Silberhorn}}]{Donohue18}%
  \BibitemOpen
  \bibfield  {author} {\bibinfo {author} {\bibfnamefont {J.~M.}\ \bibnamefont {Donohue}}, \bibinfo {author} {\bibfnamefont {V.}~\bibnamefont {Ansari}}, \bibinfo {author} {\bibfnamefont {J.}~\bibnamefont {\ifmmode \check{R}\else \v{R}\fi{}eh\'a\ifmmode~\check{c}\else \v{c}\fi{}ek}}, \bibinfo {author} {\bibfnamefont {Z.}~\bibnamefont {Hradil}}, \bibinfo {author} {\bibfnamefont {B.}~\bibnamefont {Stoklasa}}, \bibinfo {author} {\bibfnamefont {M.}~\bibnamefont {Pa\'ur}}, \bibinfo {author} {\bibfnamefont {L.~L.}\ \bibnamefont {S\'anchez-Soto}},\ and\ \bibinfo {author} {\bibfnamefont {C.}~\bibnamefont {Silberhorn}},\ }\href {https://doi.org/10.1103/PhysRevLett.121.090501} {\bibfield  {journal} {\bibinfo  {journal} {Phys. Rev. Lett.}\ }\textbf {\bibinfo {volume} {121}},\ \bibinfo {pages} {090501} (\bibinfo {year} {2018})}\BibitemShut {NoStop}%
\bibitem [{\citenamefont {Parniak}\ \emph {et~al.}(2018)\citenamefont {Parniak}, \citenamefont {Bor\'owka}, \citenamefont {Boroszko}, \citenamefont {Wasilewski}, \citenamefont {Banaszek},\ and\ \citenamefont {Demkowicz-Dobrza{\'n}ski}}]{Parniak18}%
  \BibitemOpen
  \bibfield  {author} {\bibinfo {author} {\bibfnamefont {M.}~\bibnamefont {Parniak}}, \bibinfo {author} {\bibfnamefont {S.}~\bibnamefont {Bor\'owka}}, \bibinfo {author} {\bibfnamefont {K.}~\bibnamefont {Boroszko}}, \bibinfo {author} {\bibfnamefont {W.}~\bibnamefont {Wasilewski}}, \bibinfo {author} {\bibfnamefont {K.}~\bibnamefont {Banaszek}},\ and\ \bibinfo {author} {\bibfnamefont {R.}~\bibnamefont {Demkowicz-Dobrza{\'n}ski}},\ }\href {https://doi.org/10.1103/PhysRevLett.121.250503} {\bibfield  {journal} {\bibinfo  {journal} {Phys. Rev. Lett.}\ }\textbf {\bibinfo {volume} {121}},\ \bibinfo {pages} {250503} (\bibinfo {year} {2018})}\BibitemShut {NoStop}%
\bibitem [{\citenamefont {Zhou}\ \emph {et~al.}(2019)\citenamefont {Zhou}, \citenamefont {Yang}, \citenamefont {Hassett}, \citenamefont {Rafsanjani}, \citenamefont {Mirhosseini}, \citenamefont {Vamivakas}, \citenamefont {Jordan}, \citenamefont {Shi},\ and\ \citenamefont {Boyd}}]{Zhou19}%
  \BibitemOpen
  \bibfield  {author} {\bibinfo {author} {\bibfnamefont {Y.}~\bibnamefont {Zhou}}, \bibinfo {author} {\bibfnamefont {J.}~\bibnamefont {Yang}}, \bibinfo {author} {\bibfnamefont {J.~D.}\ \bibnamefont {Hassett}}, \bibinfo {author} {\bibfnamefont {S.~M.~H.}\ \bibnamefont {Rafsanjani}}, \bibinfo {author} {\bibfnamefont {M.}~\bibnamefont {Mirhosseini}}, \bibinfo {author} {\bibfnamefont {A.~N.}\ \bibnamefont {Vamivakas}}, \bibinfo {author} {\bibfnamefont {A.~N.}\ \bibnamefont {Jordan}}, \bibinfo {author} {\bibfnamefont {Z.}~\bibnamefont {Shi}},\ and\ \bibinfo {author} {\bibfnamefont {R.~W.}\ \bibnamefont {Boyd}},\ }\href {https://doi.org/10.1364/OPTICA.6.000534} {\bibfield  {journal} {\bibinfo  {journal} {Optica}\ }\textbf {\bibinfo {volume} {6}},\ \bibinfo {pages} {534} (\bibinfo {year} {2019})}\BibitemShut {NoStop}%
\bibitem [{\citenamefont {Boucher}\ \emph {et~al.}(2020)\citenamefont {Boucher}, \citenamefont {Fabre}, \citenamefont {Labroille},\ and\ \citenamefont {Treps}}]{Boucher20}%
  \BibitemOpen
  \bibfield  {author} {\bibinfo {author} {\bibfnamefont {P.}~\bibnamefont {Boucher}}, \bibinfo {author} {\bibfnamefont {C.}~\bibnamefont {Fabre}}, \bibinfo {author} {\bibfnamefont {G.}~\bibnamefont {Labroille}},\ and\ \bibinfo {author} {\bibfnamefont {N.}~\bibnamefont {Treps}},\ }\href {https://doi.org/10.1364/OPTICA.404746} {\bibfield  {journal} {\bibinfo  {journal} {Optica}\ }\textbf {\bibinfo {volume} {7}},\ \bibinfo {pages} {1621} (\bibinfo {year} {2020})}\BibitemShut {NoStop}%
\bibitem [{\citenamefont {Rouvi\`{e}re}\ \emph {et~al.}(2024)\citenamefont {Rouvi\`{e}re}, \citenamefont {Barral}, \citenamefont {Grateau}, \citenamefont {Karuseichyk}, \citenamefont {Sorelli}, \citenamefont {Walschaers},\ and\ \citenamefont {Treps}}]{Rouviere24}%
  \BibitemOpen
  \bibfield  {author} {\bibinfo {author} {\bibfnamefont {C.}~\bibnamefont {Rouvi\`{e}re}}, \bibinfo {author} {\bibfnamefont {D.}~\bibnamefont {Barral}}, \bibinfo {author} {\bibfnamefont {A.}~\bibnamefont {Grateau}}, \bibinfo {author} {\bibfnamefont {I.}~\bibnamefont {Karuseichyk}}, \bibinfo {author} {\bibfnamefont {G.}~\bibnamefont {Sorelli}}, \bibinfo {author} {\bibfnamefont {M.}~\bibnamefont {Walschaers}},\ and\ \bibinfo {author} {\bibfnamefont {N.}~\bibnamefont {Treps}},\ }\href {https://doi.org/10.1364/OPTICA.500039} {\bibfield  {journal} {\bibinfo  {journal} {Optica}\ }\textbf {\bibinfo {volume} {11}},\ \bibinfo {pages} {166} (\bibinfo {year} {2024})}\BibitemShut {NoStop}%
\bibitem [{\citenamefont {Larson}\ and\ \citenamefont {Saleh}(2018)}]{Larson2018}%
  \BibitemOpen
  \bibfield  {author} {\bibinfo {author} {\bibfnamefont {W.}~\bibnamefont {Larson}}\ and\ \bibinfo {author} {\bibfnamefont {B.~E.~A.}\ \bibnamefont {Saleh}},\ }\href {https://doi.org/10.1364/OPTICA.5.001382} {\bibfield  {journal} {\bibinfo  {journal} {Optica}\ }\textbf {\bibinfo {volume} {5}},\ \bibinfo {pages} {1382} (\bibinfo {year} {2018})}\BibitemShut {NoStop}%
\bibitem [{\citenamefont {Tsang}\ and\ \citenamefont {Nair}(2019)}]{Tsang2019b}%
  \BibitemOpen
  \bibfield  {author} {\bibinfo {author} {\bibfnamefont {M.}~\bibnamefont {Tsang}}\ and\ \bibinfo {author} {\bibfnamefont {R.}~\bibnamefont {Nair}},\ }\href {https://doi.org/10.1364/OPTICA.6.000400} {\bibfield  {journal} {\bibinfo  {journal} {Optica}\ }\textbf {\bibinfo {volume} {6}},\ \bibinfo {pages} {400} (\bibinfo {year} {2019})}\BibitemShut {NoStop}%
\bibitem [{\citenamefont {Larson}\ and\ \citenamefont {Saleh}(2019)}]{Larson2019b}%
  \BibitemOpen
  \bibfield  {author} {\bibinfo {author} {\bibfnamefont {W.}~\bibnamefont {Larson}}\ and\ \bibinfo {author} {\bibfnamefont {B.~E.~A.}\ \bibnamefont {Saleh}},\ }\href {https://doi.org/10.1364/OPTICA.6.000402} {\bibfield  {journal} {\bibinfo  {journal} {Optica}\ }\textbf {\bibinfo {volume} {6}},\ \bibinfo {pages} {402} (\bibinfo {year} {2019})}\BibitemShut {NoStop}%
\bibitem [{\citenamefont {Hradil}\ \emph {et~al.}(2019)\citenamefont {Hradil}, \citenamefont {{\v R}eh{\'a}{\v c}ek}, \citenamefont {{S{\'a}nchez-Soto}},\ and\ \citenamefont {Englert}}]{Hradil2019}%
  \BibitemOpen
  \bibfield  {author} {\bibinfo {author} {\bibfnamefont {Z.}~\bibnamefont {Hradil}}, \bibinfo {author} {\bibfnamefont {J.}~\bibnamefont {{\v R}eh{\'a}{\v c}ek}}, \bibinfo {author} {\bibfnamefont {L.}~\bibnamefont {{S{\'a}nchez-Soto}}},\ and\ \bibinfo {author} {\bibfnamefont {B.-G.}\ \bibnamefont {Englert}},\ }\href {https://doi.org/10.1364/OPTICA.6.001437} {\bibfield  {journal} {\bibinfo  {journal} {Optica}\ }\textbf {\bibinfo {volume} {6}},\ \bibinfo {pages} {1437} (\bibinfo {year} {2019})}\BibitemShut {NoStop}%
\bibitem [{\citenamefont {Hradil}\ \emph {et~al.}(2021)\citenamefont {Hradil}, \citenamefont {Koutn\'{y}},\ and\ \citenamefont {\v{R}eh\'{a}\v{c}ek}}]{Hradil2021}%
  \BibitemOpen
  \bibfield  {author} {\bibinfo {author} {\bibfnamefont {Z.}~\bibnamefont {Hradil}}, \bibinfo {author} {\bibfnamefont {D.}~\bibnamefont {Koutn\'{y}}},\ and\ \bibinfo {author} {\bibfnamefont {J.}~\bibnamefont {\v{R}eh\'{a}\v{c}ek}},\ }\href {https://doi.org/10.1364/OL.417988} {\bibfield  {journal} {\bibinfo  {journal} {Opt. Lett.}\ }\textbf {\bibinfo {volume} {46}},\ \bibinfo {pages} {1728} (\bibinfo {year} {2021})}\BibitemShut {NoStop}%
\bibitem [{\citenamefont {Liang}\ \emph {et~al.}(2021)\citenamefont {Liang}, \citenamefont {Wadood},\ and\ \citenamefont {Vamivakas}}]{Liang2021}%
  \BibitemOpen
  \bibfield  {author} {\bibinfo {author} {\bibfnamefont {K.}~\bibnamefont {Liang}}, \bibinfo {author} {\bibfnamefont {S.~A.}\ \bibnamefont {Wadood}},\ and\ \bibinfo {author} {\bibfnamefont {A.~N.}\ \bibnamefont {Vamivakas}},\ }\href {https://doi.org/10.1103/PhysRevA.104.022220} {\bibfield  {journal} {\bibinfo  {journal} {Phys. Rev. A}\ }\textbf {\bibinfo {volume} {104}},\ \bibinfo {pages} {022220} (\bibinfo {year} {2021})}\BibitemShut {NoStop}%
\bibitem [{\citenamefont {Karuseichyk}\ \emph {et~al.}(2022)\citenamefont {Karuseichyk}, \citenamefont {Sorelli}, \citenamefont {Walschaers}, \citenamefont {Treps},\ and\ \citenamefont {Gessner}}]{Karuseichyk2022}%
  \BibitemOpen
  \bibfield  {author} {\bibinfo {author} {\bibfnamefont {I.}~\bibnamefont {Karuseichyk}}, \bibinfo {author} {\bibfnamefont {G.}~\bibnamefont {Sorelli}}, \bibinfo {author} {\bibfnamefont {M.}~\bibnamefont {Walschaers}}, \bibinfo {author} {\bibfnamefont {N.}~\bibnamefont {Treps}},\ and\ \bibinfo {author} {\bibfnamefont {M.}~\bibnamefont {Gessner}},\ }\href {https://doi.org/10.1103/PhysRevResearch.4.043010} {\bibfield  {journal} {\bibinfo  {journal} {Phys. Rev. Res.}\ }\textbf {\bibinfo {volume} {4}},\ \bibinfo {pages} {043010} (\bibinfo {year} {2022})}\BibitemShut {NoStop}%
\bibitem [{\citenamefont {Wadood}\ \emph {et~al.}(2021)\citenamefont {Wadood}, \citenamefont {Liang}, \citenamefont {Zhou}, \citenamefont {Yang}, \citenamefont {Alonso}, \citenamefont {Qian}, \citenamefont {Malhotra}, \citenamefont {S.~Hashemi~Rafsanjani}, \citenamefont {Jordan}, \citenamefont {Boyd},\ and\ \citenamefont {Vamivakas}}]{Wadood21}%
  \BibitemOpen
  \bibfield  {author} {\bibinfo {author} {\bibfnamefont {S.~A.}\ \bibnamefont {Wadood}}, \bibinfo {author} {\bibfnamefont {K.}~\bibnamefont {Liang}}, \bibinfo {author} {\bibfnamefont {Y.}~\bibnamefont {Zhou}}, \bibinfo {author} {\bibfnamefont {J.}~\bibnamefont {Yang}}, \bibinfo {author} {\bibfnamefont {A.~I.}\ \bibnamefont {Alonso}}, \bibinfo {author} {\bibfnamefont {X.-F.}\ \bibnamefont {Qian}}, \bibinfo {author} {\bibfnamefont {T.}~\bibnamefont {Malhotra}}, \bibinfo {author} {\bibfnamefont {M.}~\bibnamefont {S.~Hashemi~Rafsanjani}}, \bibinfo {author} {\bibfnamefont {A.~N.}\ \bibnamefont {Jordan}}, \bibinfo {author} {\bibfnamefont {R.~W.}\ \bibnamefont {Boyd}},\ and\ \bibinfo {author} {\bibfnamefont {A.~N.}\ \bibnamefont {Vamivakas}},\ }\href {https://doi.org/10.1364/OE.427734} {\bibfield  {journal} {\bibinfo  {journal} {Opt. Express}\ }\textbf {\bibinfo {volume} {29}},\ \bibinfo {pages} {22034} (\bibinfo {year} {2021})}\BibitemShut {NoStop}%
\bibitem [{\citenamefont {Karuseichyk}\ \emph {et~al.}(2024)\citenamefont {Karuseichyk}, \citenamefont {Sorelli}, \citenamefont {Shatokhin}, \citenamefont {Walschaers},\ and\ \citenamefont {Treps}}]{Karuseichyk2024}%
  \BibitemOpen
  \bibfield  {author} {\bibinfo {author} {\bibfnamefont {I.}~\bibnamefont {Karuseichyk}}, \bibinfo {author} {\bibfnamefont {G.}~\bibnamefont {Sorelli}}, \bibinfo {author} {\bibfnamefont {V.}~\bibnamefont {Shatokhin}}, \bibinfo {author} {\bibfnamefont {M.}~\bibnamefont {Walschaers}},\ and\ \bibinfo {author} {\bibfnamefont {N.}~\bibnamefont {Treps}},\ }\href {https://doi.org/10.1103/PhysRevA.109.043524} {\bibfield  {journal} {\bibinfo  {journal} {Phys. Rev. A}\ }\textbf {\bibinfo {volume} {109}},\ \bibinfo {pages} {043524} (\bibinfo {year} {2024})}\BibitemShut {NoStop}%
\bibitem [{\citenamefont {Sorelli}\ \emph {et~al.}(2022)\citenamefont {Sorelli}, \citenamefont {Gessner}, \citenamefont {Walschaers},\ and\ \citenamefont {Treps}}]{Sorelli2022}%
  \BibitemOpen
  \bibfield  {author} {\bibinfo {author} {\bibfnamefont {G.}~\bibnamefont {Sorelli}}, \bibinfo {author} {\bibfnamefont {M.}~\bibnamefont {Gessner}}, \bibinfo {author} {\bibfnamefont {M.}~\bibnamefont {Walschaers}},\ and\ \bibinfo {author} {\bibfnamefont {N.}~\bibnamefont {Treps}},\ }\href {https://doi.org/10.1103/PhysRevResearch.4.L032022} {\bibfield  {journal} {\bibinfo  {journal} {Phys. Rev. Res.}\ }\textbf {\bibinfo {volume} {4}},\ \bibinfo {pages} {L032022} (\bibinfo {year} {2022})}\BibitemShut {NoStop}%
\bibitem [{\citenamefont {Chrostowski}\ \emph {et~al.}(2017)\citenamefont {Chrostowski}, \citenamefont {{Demkowicz-Dobrzanski}}, \citenamefont {Jarzyna},\ and\ \citenamefont {Banaszek}}]{Chrostowski2017}%
  \BibitemOpen
  \bibfield  {author} {\bibinfo {author} {\bibfnamefont {A.}~\bibnamefont {Chrostowski}}, \bibinfo {author} {\bibfnamefont {R.}~\bibnamefont {{Demkowicz-Dobrzanski}}}, \bibinfo {author} {\bibfnamefont {M.}~\bibnamefont {Jarzyna}},\ and\ \bibinfo {author} {\bibfnamefont {K.}~\bibnamefont {Banaszek}},\ }\href {https://doi.org/10.1142/S0219749917400056} {\bibfield  {journal} {\bibinfo  {journal} {Int. J. Quantum Inform.}\ }\textbf {\bibinfo {volume} {15}},\ \bibinfo {pages} {1740005} (\bibinfo {year} {2017})}\BibitemShut {NoStop}%
\bibitem [{\citenamefont {Born}\ and\ \citenamefont {Wolf}(1999)}]{Born1999}%
  \BibitemOpen
  \bibfield  {author} {\bibinfo {author} {\bibfnamefont {M.}~\bibnamefont {Born}}\ and\ \bibinfo {author} {\bibfnamefont {E.}~\bibnamefont {Wolf}},\ }\href@noop {} {\emph {\bibinfo {title} {Principles of Optics: Electromagnetic Theory of Propagation, Interference and Diffraction of Light}}},\ \bibinfo {edition} {7th}\ ed.\ (\bibinfo  {publisher} {Cambridge University Press},\ \bibinfo {address} {Cambridge},\ \bibinfo {year} {1999})\BibitemShut {NoStop}%
\bibitem [{\citenamefont {Wilks}(1962)}]{Wilks1962}%
  \BibitemOpen
  \bibfield  {author} {\bibinfo {author} {\bibfnamefont {S.~S.}\ \bibnamefont {Wilks}},\ }\href@noop {} {\emph {\bibinfo {title} {Mathematical Statistics}}}\ (\bibinfo  {publisher} {John Wiley \& Sons, New York},\ \bibinfo {year} {1962})\BibitemShut {NoStop}%
\bibitem [{\citenamefont {Cram\'{e}r}(1999)}]{Cramer1961}%
  \BibitemOpen
  \bibfield  {author} {\bibinfo {author} {\bibfnamefont {H.}~\bibnamefont {Cram\'{e}r}},\ }\href@noop {} {\emph {\bibinfo {title} {Mathematical Methods of Statistics}}},\ Princeton paperbacks Princeton landmarks in mathematics and physics\ (\bibinfo  {publisher} {Princeton University Press, New Jersey},\ \bibinfo {year} {1999})\BibitemShut {NoStop}%
\bibitem [{\citenamefont {Fisher}\ and\ \citenamefont {Russell}(1922)}]{Fisher1922}%
  \BibitemOpen
  \bibfield  {author} {\bibinfo {author} {\bibfnamefont {R.~A.}\ \bibnamefont {Fisher}}\ and\ \bibinfo {author} {\bibfnamefont {E.~J.}\ \bibnamefont {Russell}},\ }\href {https://doi.org/10.1098/rsta.1922.0009} {\bibfield  {journal} {\bibinfo  {journal} {Philos. Trans. R. Soc. A}\ }\textbf {\bibinfo {volume} {222}},\ \bibinfo {pages} {309} (\bibinfo {year} {1922})}\BibitemShut {NoStop}%
\bibitem [{\citenamefont {Braunstein}\ and\ \citenamefont {Caves}(1994)}]{Braunstein1994}%
  \BibitemOpen
  \bibfield  {author} {\bibinfo {author} {\bibfnamefont {S.~L.}\ \bibnamefont {Braunstein}}\ and\ \bibinfo {author} {\bibfnamefont {C.~M.}\ \bibnamefont {Caves}},\ }\href {https://doi.org/10.1103/PhysRevLett.72.3439} {\bibfield  {journal} {\bibinfo  {journal} {Phys. Rev. Lett.}\ }\textbf {\bibinfo {volume} {72}},\ \bibinfo {pages} {3439} (\bibinfo {year} {1994})}\BibitemShut {NoStop}%
\bibitem [{\citenamefont {Ragy}\ \emph {et~al.}(2016)\citenamefont {Ragy}, \citenamefont {Jarzyna},\ and\ \citenamefont {Demkowicz-Dobrza{\'n}ski}}]{Ragy2016}%
  \BibitemOpen
  \bibfield  {author} {\bibinfo {author} {\bibfnamefont {S.}~\bibnamefont {Ragy}}, \bibinfo {author} {\bibfnamefont {M.}~\bibnamefont {Jarzyna}},\ and\ \bibinfo {author} {\bibfnamefont {R.}~\bibnamefont {Demkowicz-Dobrza{\'n}ski}},\ }\href {https://doi.org/10.1103/PhysRevA.94.052108} {\bibfield  {journal} {\bibinfo  {journal} {Phys. Rev. A}\ }\textbf {\bibinfo {volume} {94}},\ \bibinfo {pages} {052108} (\bibinfo {year} {2016})}\BibitemShut {NoStop}%
\bibitem [{\citenamefont {Kurdzialek}(2022)}]{Kurdzialek2022}%
  \BibitemOpen
  \bibfield  {author} {\bibinfo {author} {\bibfnamefont {S.}~\bibnamefont {Kurdzialek}},\ }\href {https://doi.org/10.22331/q-2022-04-27-697} {\bibfield  {journal} {\bibinfo  {journal} {Quantum}\ }\textbf {\bibinfo {volume} {6}},\ \bibinfo {pages} {697} (\bibinfo {year} {2022})}\BibitemShut {NoStop}%
\bibitem [{\citenamefont {Gessner}\ \emph {et~al.}(2020)\citenamefont {Gessner}, \citenamefont {Fabre},\ and\ \citenamefont {Treps}}]{Gessner20}%
  \BibitemOpen
  \bibfield  {author} {\bibinfo {author} {\bibfnamefont {M.}~\bibnamefont {Gessner}}, \bibinfo {author} {\bibfnamefont {C.}~\bibnamefont {Fabre}},\ and\ \bibinfo {author} {\bibfnamefont {N.}~\bibnamefont {Treps}},\ }\href {https://doi.org/10.1103/PhysRevLett.125.100501} {\bibfield  {journal} {\bibinfo  {journal} {Phys. Rev. Lett.}\ }\textbf {\bibinfo {volume} {125}},\ \bibinfo {pages} {100501} (\bibinfo {year} {2020})}\BibitemShut {NoStop}%
\bibitem [{\citenamefont {Grace}\ \emph {et~al.}(2020)\citenamefont {Grace}, \citenamefont {Dutton}, \citenamefont {Ashok},\ and\ \citenamefont {Guha}}]{Grace20}%
  \BibitemOpen
  \bibfield  {author} {\bibinfo {author} {\bibfnamefont {M.~R.}\ \bibnamefont {Grace}}, \bibinfo {author} {\bibfnamefont {Z.}~\bibnamefont {Dutton}}, \bibinfo {author} {\bibfnamefont {A.}~\bibnamefont {Ashok}},\ and\ \bibinfo {author} {\bibfnamefont {S.}~\bibnamefont {Guha}},\ }\href {https://doi.org/10.1364/JOSAA.392116} {\bibfield  {journal} {\bibinfo  {journal} {J. Opt. Soc. Am. A}\ }\textbf {\bibinfo {volume} {37}},\ \bibinfo {pages} {1288} (\bibinfo {year} {2020})}\BibitemShut {NoStop}%
\bibitem [{\citenamefont {de~Almeida}\ \emph {et~al.}(2021)\citenamefont {de~Almeida}, \citenamefont {Lewenstein},\ and\ \citenamefont {Skotiniotis}}]{deAlmeida21b}%
  \BibitemOpen
  \bibfield  {author} {\bibinfo {author} {\bibfnamefont {J.}~\bibnamefont {de~Almeida}}, \bibinfo {author} {\bibfnamefont {M.}~\bibnamefont {Lewenstein}},\ and\ \bibinfo {author} {\bibfnamefont {M.}~\bibnamefont {Skotiniotis}},\ }\href@noop {} {\bibfield  {journal} {\bibinfo  {journal} {arXiv preprint arXiv:2110.00986}\ } (\bibinfo {year} {2021})}\BibitemShut {NoStop}%
\bibitem [{\citenamefont {Schwartz}\ \emph {et~al.}(2013)\citenamefont {Schwartz}, \citenamefont {Levitt}, \citenamefont {Tenne}, \citenamefont {Itzhakov}, \citenamefont {Deutsch},\ and\ \citenamefont {Oron}}]{Schwartz2013}%
  \BibitemOpen
  \bibfield  {author} {\bibinfo {author} {\bibfnamefont {O.}~\bibnamefont {Schwartz}}, \bibinfo {author} {\bibfnamefont {J.~M.}\ \bibnamefont {Levitt}}, \bibinfo {author} {\bibfnamefont {R.}~\bibnamefont {Tenne}}, \bibinfo {author} {\bibfnamefont {S.}~\bibnamefont {Itzhakov}}, \bibinfo {author} {\bibfnamefont {Z.}~\bibnamefont {Deutsch}},\ and\ \bibinfo {author} {\bibfnamefont {D.}~\bibnamefont {Oron}},\ }\href {https://doi.org/10.1021/nl402552m} {\bibfield  {journal} {\bibinfo  {journal} {Nano Lett.}\ }\textbf {\bibinfo {volume} {13}},\ \bibinfo {pages} {5832} (\bibinfo {year} {2013})}\BibitemShut {NoStop}%
\bibitem [{\citenamefont {Li}\ and\ \citenamefont {Wang}(2022)}]{Li2022}%
  \BibitemOpen
  \bibfield  {author} {\bibinfo {author} {\bibfnamefont {W.}~\bibnamefont {Li}}\ and\ \bibinfo {author} {\bibfnamefont {Z.}~\bibnamefont {Wang}},\ }\href {https://doi.org/10.1364/OE.451114} {\bibfield  {journal} {\bibinfo  {journal} {Opt. Express}\ }\textbf {\bibinfo {volume} {30}},\ \bibinfo {pages} {12684} (\bibinfo {year} {2022})}\BibitemShut {NoStop}%
\bibitem [{\citenamefont {Reitzenstein}\ \emph {et~al.}(2006)\citenamefont {Reitzenstein}, \citenamefont {L\"{o}ffler}, \citenamefont {Hofmann}, \citenamefont {Kubanek}, \citenamefont {Kamp}, \citenamefont {Reithmaier}, \citenamefont {Forchel}, \citenamefont {Kulakovskii}, \citenamefont {Keldysh}, \citenamefont {Ponomarev},\ and\ \citenamefont {Reinecke}}]{Reitzenstein2006}%
  \BibitemOpen
  \bibfield  {author} {\bibinfo {author} {\bibfnamefont {S.}~\bibnamefont {Reitzenstein}}, \bibinfo {author} {\bibfnamefont {A.}~\bibnamefont {L\"{o}ffler}}, \bibinfo {author} {\bibfnamefont {C.}~\bibnamefont {Hofmann}}, \bibinfo {author} {\bibfnamefont {A.}~\bibnamefont {Kubanek}}, \bibinfo {author} {\bibfnamefont {M.}~\bibnamefont {Kamp}}, \bibinfo {author} {\bibfnamefont {J.~P.}\ \bibnamefont {Reithmaier}}, \bibinfo {author} {\bibfnamefont {A.}~\bibnamefont {Forchel}}, \bibinfo {author} {\bibfnamefont {V.~D.}\ \bibnamefont {Kulakovskii}}, \bibinfo {author} {\bibfnamefont {L.~V.}\ \bibnamefont {Keldysh}}, \bibinfo {author} {\bibfnamefont {I.~V.}\ \bibnamefont {Ponomarev}},\ and\ \bibinfo {author} {\bibfnamefont {T.~L.}\ \bibnamefont {Reinecke}},\ }\href {https://doi.org/10.1364/OL.31.001738} {\bibfield  {journal} {\bibinfo  {journal} {Opt. Lett.}\ }\textbf {\bibinfo {volume} {31}},\ \bibinfo {pages} {1738} (\bibinfo {year} {2006})}\BibitemShut {NoStop}%
\bibitem [{\citenamefont {Laussy}\ \emph {et~al.}(2008)\citenamefont {Laussy}, \citenamefont {del Valle},\ and\ \citenamefont {Tejedor}}]{Laussy2008}%
  \BibitemOpen
  \bibfield  {author} {\bibinfo {author} {\bibfnamefont {F.~P.}\ \bibnamefont {Laussy}}, \bibinfo {author} {\bibfnamefont {E.}~\bibnamefont {del Valle}},\ and\ \bibinfo {author} {\bibfnamefont {C.}~\bibnamefont {Tejedor}},\ }\href {https://doi.org/10.1103/PhysRevLett.101.083601} {\bibfield  {journal} {\bibinfo  {journal} {Phys. Rev. Lett.}\ }\textbf {\bibinfo {volume} {101}},\ \bibinfo {pages} {083601} (\bibinfo {year} {2008})}\BibitemShut {NoStop}%
\bibitem [{\citenamefont {Laucht}\ \emph {et~al.}(2009)\citenamefont {Laucht}, \citenamefont {Hauke}, \citenamefont {Villas-B\^oas}, \citenamefont {Hofbauer}, \citenamefont {B\"ohm}, \citenamefont {Kaniber},\ and\ \citenamefont {Finley}}]{Laucht2009}%
  \BibitemOpen
  \bibfield  {author} {\bibinfo {author} {\bibfnamefont {A.}~\bibnamefont {Laucht}}, \bibinfo {author} {\bibfnamefont {N.}~\bibnamefont {Hauke}}, \bibinfo {author} {\bibfnamefont {J.~M.}\ \bibnamefont {Villas-B\^oas}}, \bibinfo {author} {\bibfnamefont {F.}~\bibnamefont {Hofbauer}}, \bibinfo {author} {\bibfnamefont {G.}~\bibnamefont {B\"ohm}}, \bibinfo {author} {\bibfnamefont {M.}~\bibnamefont {Kaniber}},\ and\ \bibinfo {author} {\bibfnamefont {J.~J.}\ \bibnamefont {Finley}},\ }\href {https://doi.org/10.1103/PhysRevLett.103.087405} {\bibfield  {journal} {\bibinfo  {journal} {Phys. Rev. Lett.}\ }\textbf {\bibinfo {volume} {103}},\ \bibinfo {pages} {087405} (\bibinfo {year} {2009})}\BibitemShut {NoStop}%
\bibitem [{\citenamefont {Laussy}\ \emph {et~al.}(2009)\citenamefont {Laussy}, \citenamefont {del Valle},\ and\ \citenamefont {Tejedor}}]{Laussy2009}%
  \BibitemOpen
  \bibfield  {author} {\bibinfo {author} {\bibfnamefont {F.~P.}\ \bibnamefont {Laussy}}, \bibinfo {author} {\bibfnamefont {E.}~\bibnamefont {del Valle}},\ and\ \bibinfo {author} {\bibfnamefont {C.}~\bibnamefont {Tejedor}},\ }\href {https://doi.org/10.1103/PhysRevB.79.235325} {\bibfield  {journal} {\bibinfo  {journal} {Phys. Rev. B}\ }\textbf {\bibinfo {volume} {79}},\ \bibinfo {pages} {235325} (\bibinfo {year} {2009})}\BibitemShut {NoStop}%
\bibitem [{\citenamefont {Laucht}\ \emph {et~al.}(2010)\citenamefont {Laucht}, \citenamefont {Villas-B\^oas}, \citenamefont {Stobbe}, \citenamefont {Hauke}, \citenamefont {Hofbauer}, \citenamefont {B\"ohm}, \citenamefont {Lodahl}, \citenamefont {Amann}, \citenamefont {Kaniber},\ and\ \citenamefont {Finley}}]{Laucht2010}%
  \BibitemOpen
  \bibfield  {author} {\bibinfo {author} {\bibfnamefont {A.}~\bibnamefont {Laucht}}, \bibinfo {author} {\bibfnamefont {J.~M.}\ \bibnamefont {Villas-B\^oas}}, \bibinfo {author} {\bibfnamefont {S.}~\bibnamefont {Stobbe}}, \bibinfo {author} {\bibfnamefont {N.}~\bibnamefont {Hauke}}, \bibinfo {author} {\bibfnamefont {F.}~\bibnamefont {Hofbauer}}, \bibinfo {author} {\bibfnamefont {G.}~\bibnamefont {B\"ohm}}, \bibinfo {author} {\bibfnamefont {P.}~\bibnamefont {Lodahl}}, \bibinfo {author} {\bibfnamefont {M.-C.}\ \bibnamefont {Amann}}, \bibinfo {author} {\bibfnamefont {M.}~\bibnamefont {Kaniber}},\ and\ \bibinfo {author} {\bibfnamefont {J.~J.}\ \bibnamefont {Finley}},\ }\href {https://doi.org/10.1103/PhysRevB.82.075305} {\bibfield  {journal} {\bibinfo  {journal} {Phys. Rev. B}\ }\textbf {\bibinfo {volume} {82}},\ \bibinfo {pages} {075305} (\bibinfo {year} {2010})}\BibitemShut {NoStop}%
\bibitem [{\citenamefont {Laussy}\ \emph {et~al.}(2011)\citenamefont {Laussy}, \citenamefont {Laucht}, \citenamefont {del Valle}, \citenamefont {Finley},\ and\ \citenamefont {Villas-B\^oas}}]{Laussy2011}%
  \BibitemOpen
  \bibfield  {author} {\bibinfo {author} {\bibfnamefont {F.~P.}\ \bibnamefont {Laussy}}, \bibinfo {author} {\bibfnamefont {A.}~\bibnamefont {Laucht}}, \bibinfo {author} {\bibfnamefont {E.}~\bibnamefont {del Valle}}, \bibinfo {author} {\bibfnamefont {J.~J.}\ \bibnamefont {Finley}},\ and\ \bibinfo {author} {\bibfnamefont {J.~M.}\ \bibnamefont {Villas-B\^oas}},\ }\href {https://doi.org/10.1103/PhysRevB.84.195313} {\bibfield  {journal} {\bibinfo  {journal} {Phys. Rev. B}\ }\textbf {\bibinfo {volume} {84}},\ \bibinfo {pages} {195313} (\bibinfo {year} {2011})}\BibitemShut {NoStop}%
\bibitem [{\citenamefont {Lodahl}\ \emph {et~al.}(2015)\citenamefont {Lodahl}, \citenamefont {Mahmoodian},\ and\ \citenamefont {Stobbe}}]{Lodahl2015}%
  \BibitemOpen
  \bibfield  {author} {\bibinfo {author} {\bibfnamefont {P.}~\bibnamefont {Lodahl}}, \bibinfo {author} {\bibfnamefont {S.}~\bibnamefont {Mahmoodian}},\ and\ \bibinfo {author} {\bibfnamefont {S.}~\bibnamefont {Stobbe}},\ }\href {https://doi.org/10.1103/RevModPhys.87.347} {\bibfield  {journal} {\bibinfo  {journal} {Rev. Mod. Phys.}\ }\textbf {\bibinfo {volume} {87}},\ \bibinfo {pages} {347} (\bibinfo {year} {2015})}\BibitemShut {NoStop}%
\bibitem [{\citenamefont {Utzat}\ \emph {et~al.}(2019)\citenamefont {Utzat}, \citenamefont {Sun}, \citenamefont {Kaplan}, \citenamefont {Krieg}, \citenamefont {Ginterseder}, \citenamefont {Spokoyny}, \citenamefont {Klein}, \citenamefont {Shulenberger}, \citenamefont {Perkinson}, \citenamefont {Kovalenko},\ and\ \citenamefont {Bawendi}}]{Utzat2019}%
  \BibitemOpen
  \bibfield  {author} {\bibinfo {author} {\bibfnamefont {H.}~\bibnamefont {Utzat}}, \bibinfo {author} {\bibfnamefont {W.}~\bibnamefont {Sun}}, \bibinfo {author} {\bibfnamefont {A.~E.~K.}\ \bibnamefont {Kaplan}}, \bibinfo {author} {\bibfnamefont {F.}~\bibnamefont {Krieg}}, \bibinfo {author} {\bibfnamefont {M.}~\bibnamefont {Ginterseder}}, \bibinfo {author} {\bibfnamefont {B.}~\bibnamefont {Spokoyny}}, \bibinfo {author} {\bibfnamefont {N.~D.}\ \bibnamefont {Klein}}, \bibinfo {author} {\bibfnamefont {K.~E.}\ \bibnamefont {Shulenberger}}, \bibinfo {author} {\bibfnamefont {C.~F.}\ \bibnamefont {Perkinson}}, \bibinfo {author} {\bibfnamefont {M.~V.}\ \bibnamefont {Kovalenko}},\ and\ \bibinfo {author} {\bibfnamefont {M.~G.}\ \bibnamefont {Bawendi}},\ }\href {https://doi.org/10.1126/science.aau7392} {\bibfield  {journal} {\bibinfo  {journal} {Science}\ }\textbf {\bibinfo {volume} {363}},\ \bibinfo {pages} {1068} (\bibinfo {year} {2019})}\BibitemShut {NoStop}%
\bibitem [{\citenamefont {Liu}\ \emph {et~al.}()\citenamefont {Liu}, \citenamefont {Li}, \citenamefont {Liu}, \citenamefont {Qiu}, \citenamefont {Ma}, \citenamefont {Nie}, \citenamefont {Meng}, \citenamefont {Hu}, \citenamefont {Ni}, \citenamefont {Niu}, \citenamefont {Qiu}, \citenamefont {Wang},\ and\ \citenamefont {Liu}}]{Liu2024}%
  \BibitemOpen
  \bibfield  {author} {\bibinfo {author} {\bibfnamefont {S.}~\bibnamefont {Liu}}, \bibinfo {author} {\bibfnamefont {X.}~\bibnamefont {Li}}, \bibinfo {author} {\bibfnamefont {H.}~\bibnamefont {Liu}}, \bibinfo {author} {\bibfnamefont {G.}~\bibnamefont {Qiu}}, \bibinfo {author} {\bibfnamefont {J.}~\bibnamefont {Ma}}, \bibinfo {author} {\bibfnamefont {L.}~\bibnamefont {Nie}}, \bibinfo {author} {\bibfnamefont {Y.}~\bibnamefont {Meng}}, \bibinfo {author} {\bibfnamefont {X.}~\bibnamefont {Hu}}, \bibinfo {author} {\bibfnamefont {H.}~\bibnamefont {Ni}}, \bibinfo {author} {\bibfnamefont {Z.}~\bibnamefont {Niu}}, \bibinfo {author} {\bibfnamefont {C.-W.}\ \bibnamefont {Qiu}}, \bibinfo {author} {\bibfnamefont {X.}~\bibnamefont {Wang}},\ and\ \bibinfo {author} {\bibfnamefont {J.}~\bibnamefont {Liu}},\ }\href {https://doi.org/10.1038/s41566-024-01449-4} {\bibfield  {journal} {\bibinfo  {journal} {Nat. Photonics}\ }\textbf {\bibinfo {volume} {18}},\ \bibinfo {pages} {967}}\BibitemShut {NoStop}%
\bibitem [{\citenamefont {Vašinka}\ \emph {et~al.}()\citenamefont {Vašinka}, \citenamefont {Lee}, \citenamefont {Stalker}, \citenamefont {Mitryakhin}, \citenamefont {Solovev}, \citenamefont {Stephan}, \citenamefont {Höfling}, \citenamefont {Eilenberger}, \citenamefont {Tongay}, \citenamefont {Schneider}, \citenamefont {Ježek},\ and\ \citenamefont {Predojević}}]{Vasinka2025}%
  \BibitemOpen
  \bibfield  {author} {\bibinfo {author} {\bibfnamefont {D.}~\bibnamefont {Vašinka}}, \bibinfo {author} {\bibfnamefont {J.}~\bibnamefont {Lee}}, \bibinfo {author} {\bibfnamefont {C.}~\bibnamefont {Stalker}}, \bibinfo {author} {\bibfnamefont {V.}~\bibnamefont {Mitryakhin}}, \bibinfo {author} {\bibfnamefont {I.}~\bibnamefont {Solovev}}, \bibinfo {author} {\bibfnamefont {S.}~\bibnamefont {Stephan}}, \bibinfo {author} {\bibfnamefont {S.}~\bibnamefont {Höfling}}, \bibinfo {author} {\bibfnamefont {F.}~\bibnamefont {Eilenberger}}, \bibinfo {author} {\bibfnamefont {S.~A.}\ \bibnamefont {Tongay}}, \bibinfo {author} {\bibfnamefont {C.}~\bibnamefont {Schneider}}, \bibinfo {author} {\bibfnamefont {M.}~\bibnamefont {Ježek}},\ and\ \bibinfo {author} {\bibfnamefont {A.}~\bibnamefont {Predojević}},\ }\href {https://doi.org/10.48550/arXiv.2510.06076} {\bibinfo {title} {Universal super-resolution framework for imaging of quantum dots}},\ \Eprint {https://arxiv.org/abs/2510.06076} {2510.06076} \BibitemShut {NoStop}%
\bibitem [{\citenamefont {Huang}\ \emph {et~al.}()\citenamefont {Huang}, \citenamefont {Miranda}, \citenamefont {Liu}, \citenamefont {Cheng}, \citenamefont {Dwir}, \citenamefont {Rudra}, \citenamefont {Chang}, \citenamefont {Kapon},\ and\ \citenamefont {Wong}}]{Huang2025}%
  \BibitemOpen
  \bibfield  {author} {\bibinfo {author} {\bibfnamefont {J.}~\bibnamefont {Huang}}, \bibinfo {author} {\bibfnamefont {A.}~\bibnamefont {Miranda}}, \bibinfo {author} {\bibfnamefont {W.}~\bibnamefont {Liu}}, \bibinfo {author} {\bibfnamefont {X.}~\bibnamefont {Cheng}}, \bibinfo {author} {\bibfnamefont {B.}~\bibnamefont {Dwir}}, \bibinfo {author} {\bibfnamefont {A.}~\bibnamefont {Rudra}}, \bibinfo {author} {\bibfnamefont {K.-C.}\ \bibnamefont {Chang}}, \bibinfo {author} {\bibfnamefont {E.}~\bibnamefont {Kapon}},\ and\ \bibinfo {author} {\bibfnamefont {C.~W.}\ \bibnamefont {Wong}},\ }\href {https://doi.org/10.1038/s42005-025-02051-y} {\bibfield  {journal} {\bibinfo  {journal} {Commun. Phys.}\ }\textbf {\bibinfo {volume} {8}},\ \bibinfo {pages} {152}},\ \Eprint {https://arxiv.org/abs/40224499} {40224499} \BibitemShut {NoStop}%
\end{thebibliography}%
\end{document}